\documentclass{aa}  

\usepackage{footnote}
\usepackage{graphicx}
\usepackage{hyperref}
\hypersetup{
    colorlinks=true,                
    citecolor=blue,              
}
\usepackage[switch]{lineno} 
\usepackage{appendix}
\usepackage{color}
\usepackage{natbib}
\usepackage{txfonts}
\newcommand{\code}[1]{\texttt{#1}\xspace}
\newcommand{\teff}{\ensuremath{T_\mathrm{eff}}\xspace}
\newcommand{\logg}{\ensuremath{\log\,g}\xspace}
\newcommand{\Gaia}{\gaia}
\newcommand{\gaia}{\textit{Gaia}\xspace}
\newcommand{\unit}[1]{\ensuremath{\mathrm{\,#1}}\xspace}
\newcommand{\feh}{\unit{[Fe/H]}}

\newcommand{\msun}{\unit{M_\odot}}

\newcommand{\Angstrom}{\,{\AA}}

\usepackage{color}

\begin{document}

   \title{The \emph{R}-Process Alliance: Hunting for gold in the near-UV spectrum of 2MASS~J05383296--5904280
\thanks{Based on observations made with the NASA/ESA Hubble Space Telescope, obtained at the Space Telescope Science Institute (STScI), which is operated by the Association of Universities for Research in Astronomy, Inc.\ (AURA) under NASA contract NAS~5-26555.
These observations are primarily associated with program GO-15951. This paper also includes data gathered with the 6.5~meter Magellan Telescopes located at Las Campanas Observatory, Chile.}}
   
\titlerunning{}
\authorrunning{Hansen et al.}


   \author{Terese T. Hansen
          \inst{1}
          \and
          Ian U. Roederer\inst{2,3}
          \and
          Shivani P. Shah\inst{4}
          \and
          Rana Ezzeddine\inst{4,3}
          \and
          Timothy C. Beers\inst{5,3}
          \and
          Anna Frebel\inst{6,3}
          \and
          Erika M. Holmbeck\inst{7,3}
          \and
          Vinicius M.\ Placco\inst{8}
          \and
          Charli M. Sakari\inst{9}
          \and
          Alexander Ji\inst{10}
          \and
          Jennifer L. Marshall\inst{11}
          \and
          Mohammad K. Mardini\inst{12,13,6,3}
          \and
          Anirudh Chiti\inst{10,14}
                  }
   \institute{Department of Astronomy, Stockholm University, AlbaNova
University Center, SE-106 91 Stockholm, Sweden \email{thidemannhansen@gmail.com}
         \and
         Department of Physics, North Carolina State University, Raleigh, NC 27695, USA
         \and
         Joint Institute for Nuclear Astrophysics -- Center for the Evolution of the Elements (JINA-CEE), USA
         \and
         Department of Astronomy, University of Florida, Bryant Space Science Center, Gainesville, FL 32611, USA
         \and
         Department of Physics and Astronomy, University of Notre Dame, Notre Dame, IN 46556, USA
         \and
        Department of Physics and Kavli Institute for Astrophysics and Space Research, Massachusetts Institute of Technology, 
Cambridge, MA 02139, USA
\and
Lawrence Livermore National Laboratory, 7000 East Avenue, Livermore, CA 94550, USA
\and
NSF NOIRLab, Tucson, AZ 85719, USA
\and
Department of Physics and Astronomy, San Francisco State University,
San Francisco, CA 94132, USA
\and
Department of Astronomy \& Astrophysics, University of Chicago, 5640 S Ellis Avenue, Chicago, IL 60637, USA
\and
Mitchell Institute for Fundamental Physics and Astronomy and Department of Physics and Astronomy, Texas A\&M University, College Station, TX 77843-4242, USA
\and
Department of Physics, Zarqa University, Zarqa 13110, Jordan
\and
Jordanian Astronomical Virtual Observatory, Zarqa University, Zarqa 13110, Jordan
\and
Kavli Institute for Cosmological Physics, University of Chicago, Chicago, IL 60637, USA
             }



 
  \abstract
   {Over the past few years, the \emph{R}-Process Alliance (RPA) has successfully carried out a search for stars that are highly enhanced in elements produced via the rapid neutron-capture (\emph{r}-) process. In particular, the RPA has identified a number of relatively bright, highly \emph{r}-process-enhanced (\emph{r}-II) stars, suitable for observations with the Hubble Space Telescope (HST), facilitating abundance derivation of elements such as gold (Au) and cadmium (Cd).}
   {This paper presents the detailed abundances derived for the metal-poor ($\mathrm{[Fe/H]} = -2.55$) highly \emph{r}-process-enhanced ($\mathrm{[Eu/Fe]} = +1.29$) \emph{r}-II star 2MASS~J05383296--5904280. }
   {1D LTE elemental abundances are derived via equivalent width and spectral synthesis using high-resolution high signal-to-noise near-UV HST/STIS and optical Magellan/MIKE spectra.}
   {Abundances are determined for 43 elements, including 26 neutron-capture elements. In particular, abundances of the rarely studied elements Nb, Mo, Cd, Lu, Os, Pt, and Au are derived from the HST spectrum. These results, combined with RPA near-UV observations of two additional \emph{r}-II stars, increase the number of Cd abundances derived for \emph{r}-process-enriched stars from seven to ten and Au abundances from four to seven. A large star-to-star scatter is detected for both of these elements, highlighting the need for more detections enabling further investigations, specifically into possible non-LTE (local thermodynamical equilibrium) effects.}
   {}
   

   \maketitle

%

\section{Introduction \label{sec:intro}}
The origin and chemical evolution of elements such as gold and silver have long been investigated in nuclear astrophysics. These elements, along with half of the other isotopes heavier than iron, are produced via the rapid-neutron capture process or \emph{r}-process. This process was predicted theoretically almost 70 years ago \citep{burbidge1957,cameron1957}, but the astrophysical site(s) associated with it are still heavily debated. For decades, neutron star mergers (NSM) were strong candidates for at least one astrophysical \emph{r}-process site \citep{lattimer1974}. This was confirmed in 2017 when the Laser Interferometer Gravitational-Wave (LIGO) observatory detected a gravitational wave signal from an NSM \citep{abbott2017}. Extensive photometric and spectroscopic observations of the corresponding kilonova (AT2017gfo/SSS17a) revealed a signal consistent with the decay of heavy-element isotopes produced in the merger \citep{drout2017}. Potentially, 3--13 Earth masses of gold were produced in this event; however, there are large uncertainties in the assumed abundance composition resulting from this one event \citep{cote2018}. Furthermore, galactic chemical-evolution models and the contents of neutron-capture elements in the oldest stars in the Milky Way (MW) halo and its satellites suggest that NSMs may not be the only site of \emph{r}-process-element production \citep{cote2017}. Since the discovery of AT2017gfo, a renewed interest in the \emph{r}-process has resulted in a considerable effort from theorists to identify additional heavy-element production sites. Currently, in addition to NSMs and neutron star--black hole mergers \citep{lattimer1974,surman2008,wehmeyer2019}, various types of supernovae (SNe), such as collapsars \citep{siegel2019}, magneto-rotationally driven SNe \citep{fujimoto2008}, and common-envelope jet SNe \citep{grichener2019} have also been suggested, and most recently magnetar giant flares have been added to the list of possible $r$-process element production sites \citep{patel2025}. 

Since direct observations of \emph{r}-process-element production sites such as NSMs and exotic SNe are still very sparse, and new observations will be difficult to obtain as these events are rare (and often also faint), we need to look in other places for clues concerning the details of the \emph{r}-process. One such probe is the abundances of the oldest and most metal-poor stars in the MW. The chemical compositions of these stars, mapped through detailed abundance analyses, provide a direct fingerprint of the elements produced by the stellar generation(s) prior to their birth. 

In 2016, The \emph{R}-Process Alliance (RPA) initiated a successful search for \emph{r}-process-enriched stars to investigate the \emph{r}-process. In the first five data release papers \citep{hansen2018,sakari2018a,holmbeck2020,ezzeddine2020,bandyopadhyay2024} the RPA has identified over 70 highly \emph{r}-process-enhanced \emph{r}-II stars ($\mathrm{[Eu/Fe]} > +0.7$; \citealt{holmbeck2020}) and several hundred moderately enhanced \emph{r}-I stars ($+0.3 < \mathrm{[Eu/Fe]} \leq 0.7$; \citealt{holmbeck2020}). For several of these stars, the RPA has been able to derive extremely detailed \emph{r}-process-element abundance patterns, including more than 20 neutron-capture elements, e.g., for RAVE~J203843.2--002333 \citep{placco2017}, RAVE~J153830.9--180424 \citep{sakari2018b}, HD~222925 \citep{roederer2018a,roederer2022a}, 2MASS~J09544277+5246414 \citep{holmbeck2018}, 2MASS~J14325334--4125494 \citep{cain2018}, RAVE~J094921.8--161722 \citep{gull2018}, RAVE~J183013.5--455510 \citep{placco2020}, 2MASS~J1521399--3538094 \citep{cain2020}, 2MASS~J22132050--5137385 \citep{roederer2024}, and 2MASS~J00512646--1053170 \citep{shah2024}, revealing new details on the \emph{r}-process. 

However, abundances for a number of neutron-capture elements, such as cadmium (Cd) and gold (Au), cannot be obtained from ground-based observations, as these elements only have absorption features in the near-ultraviolet (near-UV) region of the electromagnetic spectrum around 2000 to 3200\Angstrom, for the conditions found in late-type stellar atmospheres and the typical abundances of these elements. Elements like Cd and Au could enable critical constraints on the nuclear properties of \emph{r}-process nucleosynthesis. For instance, a recent result from the RPA proposed that select elements can be enhanced by fission-fragment deposition from transuranic elements \citep{roederer2023}. However, the exact extent of elements affected by this deposition is less clear due to the limited number of abundances for elements like Cd and Au and the unconstrained, theoretical yields of transuranic nuclei. Hence, part of the RPA survey effort has also been focused on identifying relatively bright stars for which near-UV spectra can be obtained with the near-UV-sensitive Hubble Space Telescope (HST) to investigate the behaviour of these additional neutron-capture elements in \emph{r}-process-enriched stars. 

Here, we present a detailed abundance analysis of the third RPA star observed in the near-UV with the HST, namely the warm horizontal-branch star 2MASS~J05383296--5904280 (hereafter J0538), with $\mathrm{[Fe/H]} = -2.55$. The other two RPA stars with near-UV HST spectra are 2MASS~J00512646--1053170, for which \citet{shah2024} derived the highest Au enhancement found in an \emph{r}-II star to date, and HD~222925, for which \citet{roederer2022a} derived the most complete abundance pattern of any object outside the Solar System, including 42 neutron-capture elements. Abundances of Au have been derived for all three stars, almost doubling the number of Au abundances available for \emph{r}-process-enriched stars in the literature \citep{cowan2002,sneden2003,barbuy2011,roederer2012}, allowing us for the first time to investigate the behaviour of this element more systematically.

This paper is organized as follows. Section \ref{sec:obs} describes the optical and near-UV observations of J0538. The analysis is described in Section \ref{sec:analysis}. Section \ref{sec:results} presents our results, which are discussed in Section \ref{sec:discuss}. Section \ref{sec:summary} provides a summary.

\section{Observations \label{sec:obs}}
\subsection{Optical spectrum}
Basic properties for J0538, including right ascension (R.A.), declination (Decl.), and magnitudes, are listed in Table \ref{tab:obs}. The star was first observed as part of the RPA ``snapshot'' ($R \sim 30,000$ and signal-to-noise (SNR) $\sim$30 at 4100\Angstrom) survey \citep{hansen2018} on November 24, 2017, with the echelle spectrograph on the du Pont telescope \citep{bowen1973}. The analysis was published in the fourth RPA data release \citep{holmbeck2020}, which determined a Eu abundance of $\mathrm{[Eu/Fe]} = +1.28$, and a $\mathrm{[Ba/Eu]} = -0.52$, classifying it as an \emph{r}-II star. A high-resolution, high SNR ``portrait'' spectrum was subsequently obtained with the MIKE spectrograph \citep{bernstein2003} on November 15, 2018. Two 1200-s exposures were obtained using the $0\farcs5 \times 5\farcs0$ slit and 2$\times$1 binning, yielding a resolving power of $R\sim 55,000$ in the blue and $R\sim 45,000$ in the red, with a spectral coverage of 3330--5060\Angstrom\ and 4850--9400\Angstrom\ in the blue and red channels, respectively. The data were reduced using the CarPy software \citep{kelson2003}; the final spectrum has a SNR of 335 pix$^{-1}$ at 4500\Angstrom.

\subsection{Near-UV spectrum}
A near-UV spectrum of J0538 was obtained with HST/STIS \citep{kimble1998,woodgate1998} on April 16 and 21, 2020 \citep[][GO-15951]{hansen2019}. The observations were carried out using the E230M echelle grating centred at 2707\Angstrom, providing a wavelength coverage of $\sim$2275--3119\Angstrom. The $0\farcs2 \times 0\farcs06$ slit was employed, yielding a resolving power of $R\sim 30,000$. The star was observed over five orbits, with three orbits in the first visit and two in the second. The total exposure time over the five orbits was 3.63~hours. The spectra were processed automatically by the CALSTIS software package and downloaded from the Mikulski Archive for Space Telescopes. Each of the five spectra was cross-correlated against a spectrum of the star HD~84937 ($V_{\text{helio}}= -15.05$~km\,s$^{-1}$; \citealt{gaiarv}), obtained with the same setup \citep[][GO-7402]{peterson2001}, and then radial velocity-shifted to rest. The shifted spectra were then normalized and co-added, resulting in an SNR of 44 at 2707\Angstrom\ in the final spectrum.

\begin{table*}
\caption{Properties of 2MASS~J05383296--5904280. \label{tab:obs}}
\label{tab:star}
    \centering
    \begin{tabular}{llrll}
    \hline\hline
    Quantity & Symbol & Value & Unit & Ref\\
    \hline
R.A.  & $\alpha$(J2000) &  05:38:32.96 & hh:mm:ss.ss & Simbad\\
Decl. & $\delta$(J2000) & $-$59:04:28.11 & hh:mm:ss.ss & Simbad\\
Galactic longitude & $l$ & 267.7 & degrees & Simbad \\
Galactic latitude  & $b$ & $-$32.3 & degrees & Simbad \\
Parallax & $\varpi$ & 0.8804$\pm$0.0109 & mas & \citet{gaiadr3}\\
Distance & $d$ & 1128$^{+14}_{-15}$ & pc & \citet{bailorjones2021} \\ 
Mass  & Mass &  0.80$\pm$0.08 & $M_\odot$ & Assumed \\
$V$ magnitude   & $V$   & 9.887$\pm$0.008  & mag & \citet{munari2014}\\
$G$ magnitude   & $G$   & 9.602$\pm$0.003	& mag & \citet{gaiadr3}\\
$BP$ magnitude  & $BP$  & 9.914$\pm$0.088 & mag & \citet{gaiadr3}\\
$RP$ magnitude  & $RP$  & 9.10$\pm$0.052 & mag & \citet{gaiadr3}\\
$K$ magnitude   & $K$   &  8.122 $\pm$0.023 & mag & \citet{cutri2003}\\
Color excess & $E(B-V)$ &  0.056$\pm$0.001 & mag & \citet{schlafly2011}\\
Effective Temperature & $T_{\rm eff}$  & 5819$\pm$57 & K & This work \\
Log of surface gravity & $\log g$ & 2.09$\pm$0.08 & (cgs) & This work \\
Microturbulent velocity & $\xi$ & 2.60$\pm$0.05 & km~s$^{-1}$ & This work\\
Model metallicity & $\mathrm{[M/H]}$ & $-$2.55$\pm$0.07 &  & This work\\
\hline
    \end{tabular}
\end{table*}

\subsection{Radial velocities}
Radial velocities determined for J0538 from the snapshot and portrait spectra are listed in Table \ref{tab:rv}; a clear radial-velocity variation is detected between the two observations, suggesting the star is part of a binary system. We, therefore, obtained additional spectra of the star with the MIKE spectrograph for radial-velocity measurements using the $0\farcs7 \times 0\farcs5$ slit and 2$\times$2 binning. For each spectrum, we determine a radial velocity via order-by-order cross-correlation with a spectrum of the radial velocity standard star HD~122563 ($V_{\text{helio}}$= $-$26.13; \citealt{gaiarv}), obtained with the same setting as the target star. Final velocities are listed in Table \ref{tab:rv}, along with the standard deviations and number of echelle orders used. The reported velocity is the mean of the individual velocities measured in the employed echelle orders, and $\sigma$ is their standard deviation. Unfortunately, we are not able to compute an orbit for the system with the current set of radial-velocity measurements.

\begin{table}[]
\caption{Heliocentric radial velocities measured for J0538.}
\label{tab:rv}
    \centering    
    \begin{tabular}{lrrr}
    \hline\hline
MJD & $V_{\text{helio}}$ & $\sigma$ & \# orders \\
  & km s$^{-1}$ &  km s$^{-1}$ \\
  \hline
58081.28257$^*$ & 189.09 & 0.37 &  \\  
58437.51464$^+$ & 194.74 & 0.73 & 43 \\  
59641.51270     & 197.30 & 0.51 & 41 \\  
59972.52637     & 197.92 & 0.61 & 42 \\
60553.92157     & 198.17 & 0.67 & 44 \\  
60653.65973     & 197.70 & 0.51 & 46 \\
\hline
    \end{tabular}
    \tablefoot{$^*$ Snapshot spectrum, $^+$ portrait spectrum. The velocity of the snapshot spectrum is taken from \citet{holmbeck2020}. Therefore, no \# orders value is provided.}
\end{table}

\section{Stellar parameter determination and abundance analysis\label{sec:analysis}} 
The stellar parameters were determined following the standard RPA procedure (see, e.g., \citealt{roederer2022a,shah2024}). Effective temperatures \teff is derived from the Gaia $G$, $BP$, and $RP$ bands and 2MASS $K$ magnitudes using the colour-temperature relations from \citet{mucciarelli2021}. The \Gaia magnitudes were de-reddened following \citet{Gaiared}, using the $E(B-V)$ value listed in Table \ref{tab:obs} obtained from the \cite{schlafly2011} dust maps. To de-redden the $K$ magnitude, we use the extinction coefficient from \cite{mccall2004}. Following the $T_{\rm eff}$ determination, we determine \logg from the fundamental relation:
\begin{eqnarray}
\log g = 4 \log \teff + \log(M/\msun) - 10.61 + 0.4\cdot(BC_{V}
\nonumber\\
  + m_{V} - 5\log (d) + 5 - 3.1 \cdot E(B-V) - M_{\rm bol,\odot}).
\end{eqnarray}
Here, $M$ is the mass of the star, assumed to be 0.8 $\pm$ 0.08 \msun, $BC_V$ is the bolometric correction in the $V$ band \citep{casagrande2014}, $m_V$ is the apparent $V$ magnitude, $d$ is the distance in pc from \citet{bailorjones2021}, and $M_{\rm bol,\odot}$ is the Solar bolometric magnitude, 4.75. The constant 10.61 is calculated from the Solar constants: $\log(T_{\mathrm{eff}})_{\odot} = 3.7617$ and $\log g_{\odot} = 4.438$.
Finally, the model-atmosphere metallicity $\mathrm{[M/H]}$ and microturbulence $\xi$ are determined from equivalent width (EW) measurements of \ion{Fe}{i} and \ion{Fe}{ii} lines, using the [\ion{Fe}{ii}/H] abundance as the model atmosphere metallicity. The final stellar parameters are listed in Table \ref{tab:obs}. The uncertainty on the photometric \teff is determined as the 1-$\sigma$ standard deviation for Monte Carlo resamples of the input parameter values and similar for \logg using the \teff input. The corresponding effects on \feh and $\xi$ are determined by varying \teff and \logg accordingly and redetermining \feh and $\xi$, which results in the following uncertainties: $\delta$\teff, $\delta$\logg, $\delta$\feh, $\delta\xi$ = 47\,K, 0.07\,dex, 0.04\,dex, and 0.1 km s$^{-1}$, respectively. We also determine the uncertainty on the stellar parameters arising from the scatter in the individual \ion{Fe}{i} and \ion{Fe}{ii} line abundances used, resulting in $\delta$\teff, $\delta$\logg, $\delta$ \feh, $\delta\xi$ = 36\,K, 0.03\,dex, 0.06\,dex, and 0.17~km~s$^{-1}$, respectively. To reach the final parameter uncertainties listed in Table \ref{tab:obs}, we add these values in quadrature.  

Abundances for J0538 have been derived via EW analysis and spectral synthesis using the analysis code 
\code{SMHr}\footnote{\href{https://github.com/andycasey/smhr}{https://github.com/andycasey/smhr}} to run the 1D radiative transfer code \code{MOOG}\footnote{\href{https://github.com/alexji/moog17scat}{https://github.com/alexji/moog17scat}} \citep{sneden1973,sobeck2011}, assuming local thermodynamical equilibrium (LTE). We use $\alpha$-enhanced ($\mathrm{[\alpha/Fe] = +0.4}$) \code{ATLAS9} model atmospheres \citep{castelli2003}, and line lists generated from \code{linemake}\footnote{\href{https://github.com/vmplacco/linemake}{https://github.com/vmplacco/linemake}} \citep{placco2021}; updates following \citet{roederer2018a,roederer2022a} as input, and Solar abundances were taken from \citet{asplund2009}. The synthesis also includes isotopic shifts and hyperfine-structure (HFS) broadening, where applicable, employing the \emph{r}-process isotopic ratios from \citet{sneden2008}. Table \ref{tab:lines} in the appendix lists the atomic data for the lines used in the analysis, along with the measured EW and derived abundance of each line.

\begin{figure*}[hbt!]
\centering
\includegraphics[width=\linewidth]{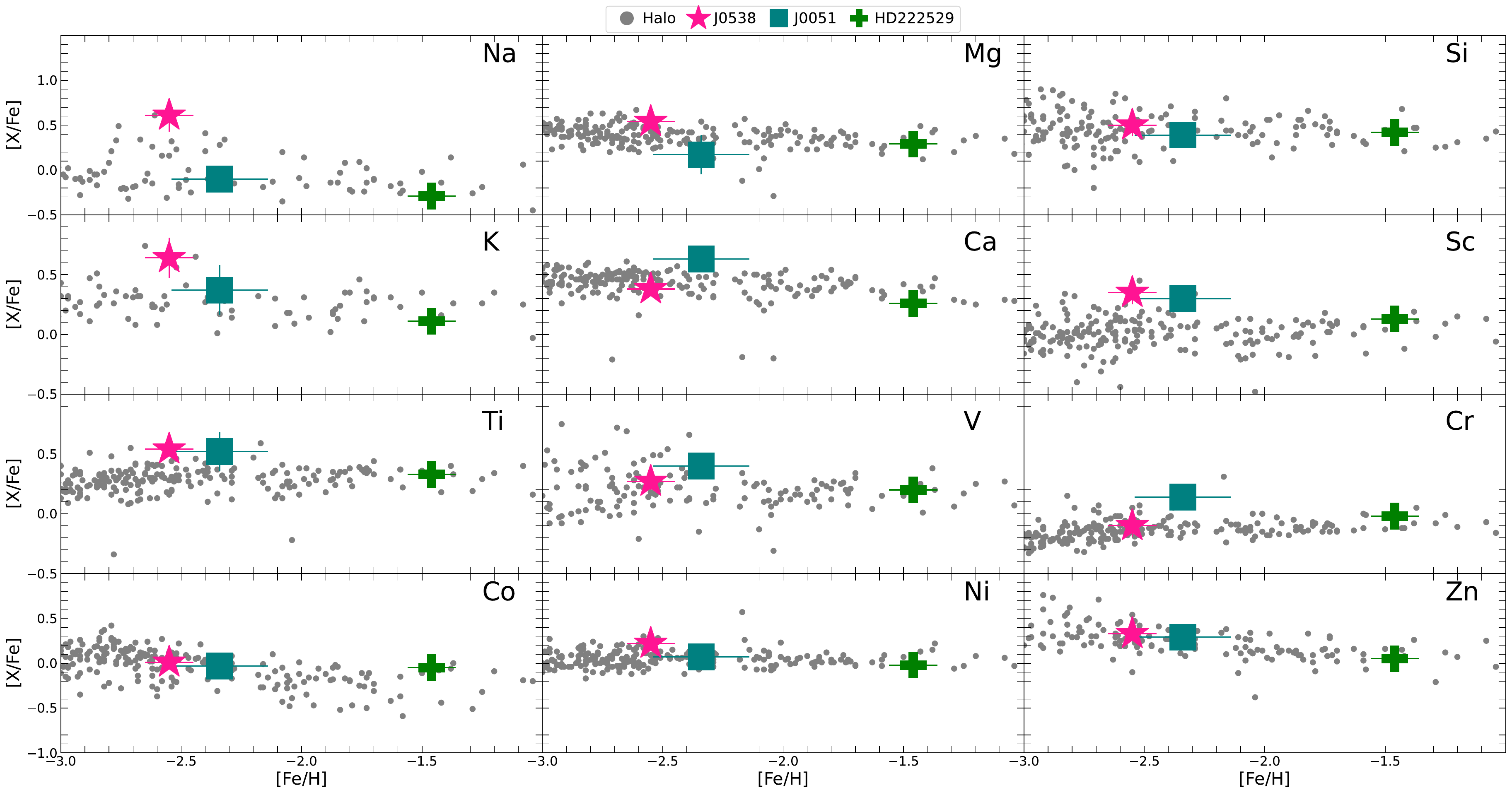}
\caption{Derived $\mathrm{[X/Fe]}$ abundances for the three RPA stars observed with the HST: J0538 (pink star, this work), HD~222925 (green cross, \citealt{roederer2022a}), and 2MASS~J00512646--1053170 (blue square,
\citealt{shah2024}), compared to stellar abundances from the MW halo \citep[grey dots;][]{roederer2014a}. \label{fig:abun_tot}}
\end{figure*}

\section{Results\label{sec:results}}
We derive abundances for 43 elements across 51 species. Table \ref{tab:abunAll} lists the final 1D and LTE abundances and associated uncertainties. For 30 species, we identified and analysed lines in the near-UV spectrum. Abundances derived for these elements are listed explicitly in Table \ref{tab:abunUV}.  Below, we highlight possible corrections to these abundances due to departure from the LTE assumption employed in this work. However, we leave it up to the reader to decide which corrections to employ when using the abundances listed in Table \ref{tab:abunAll} and \ref{tab:abunUV}. The abundance uncertainties are determined by propagating through the stellar-parameter uncertainties, following the procedure outlined in \cite{ji2020b}. This method performs a line-by-line analysis and includes both statistical and systematic uncertainties. In addition to stellar-parameter uncertainties, we add in quadrature a 0.2\,dex uncertainty to all abundances derived from two or fewer lines and 0.1\,dex to all abundances to account for continuum placement, atomic data uncertainties, and other systematic uncertainties. A summary of all the uncertainties calculated for each species is listed in Table \ref{tab:err}.

\begin{table}
\caption{Abundance summary for J0538.}
\label{tab:abunAll}
\resizebox{\columnwidth}{!}{
    \centering
    \begin{tabular}{lrrrrrr}
    \hline\hline
 El. & N & $\log\epsilon$ (X) & $\mathrm{[X/H]}$ & $\sigma_{\mathrm{[X/H]}}$ & $\mathrm{[X/Fe]}$ & $\sigma_{\mathrm{[X/Fe]}}$ \\
 &  &  &  & [dex] & & [dex]\\
\hline
\ion{CH}{}   &     1 & $<$4.88 &$<-$3.55 &\dots &$<-$0.96 & \dots \\
\ion{NH}{}   &     1 & $<$5.93 &$<-$1.90 &\dots &$<+$0.70 & \dots \\
\ion{O}{i}   &     3 &    7.13 & $-$1.56 & 0.10 & $+$1.03 & 0.11 \\
\ion{Na}{i}  &     2 &    4.54 & $-$1.70 & 0.25 & $+$0.90 & 0.25 \\
\ion{Mg}{i}  &    12 &    5.47 & $-$2.13 & 0.11 & $+$0.46 & 0.11 \\
\ion{Al}{i}  &     5 &    3.28 & $-$3.17 & 0.20 & $-$0.57 & 0.19 \\
\ion{Al}{ii} &     1 &    3.87 & $-$2.58 & 0.23 & $+$0.02 & 0.23 \\
\ion{Si}{i}  &     4 &    5.22 & $-$2.29 & 0.14 & $+$0.30 & 0.14 \\
\ion{Si}{ii} &     2 &    5.67 & $-$1.84 & 0.25 & $+$0.75 & 0.25 \\
\ion{P}{i}   &     1 &    2.69 & $-$2.72 & 0.26 & $-$0.13 & 0.26 \\
\ion{K}{i}   &     2 &    3.28 & $-$1.75 & 0.23 & $+$0.84 & 0.23 \\
\ion{Ca}{i}  &    19 &    4.14 & $-$2.20 & 0.11 & $+$0.40 & 0.11 \\
\ion{Sc}{ii} &    12 &    0.83 & $-$2.32 & 0.11 & $+$0.27 & 0.12 \\
\ion{Ti}{i}  &    16 &    2.80 & $-$2.15 & 0.11 & $+$0.44 & 0.11 \\
\ion{Ti}{ii} &    36 &    2.82 & $-$2.13 & 0.11 & $+$0.47 & 0.11 \\
\ion{V}{ii}  &    23 &    1.59 & $-$2.34 & 0.11 & $+$0.25 & 0.11 \\
\ion{Cr}{i}  &    11 &    2.93 & $-$2.71 & 0.11 & $-$0.12 & 0.11 \\
\ion{Cr}{ii} &    15 &    3.24 & $-$2.40 & 0.11 & $+$0.19 & 0.11 \\
\ion{Mn}{i}  &     7 &    2.32 & $-$3.11 & 0.12 & $-$0.52 & 0.12 \\
\ion{Mn}{ii} &     5 &    2.56 & $-$2.87 & 0.11 & $-$0.28 & 0.12 \\
\ion{Fe}{i}  &   115 &    4.91 & $-$2.59 & 0.10 & \dots   &\dots \\
\ion{Fe}{ii} &    13 &    4.98 & $-$2.52 & 0.11 & $+$0.08 &\dots \\
\ion{Co}{i}  &    18 &    2.49 & $-$2.50 & 0.11 & $+$0.09 & 0.11 \\
\ion{Co}{ii} &     6 &    2.48 & $-$2.51 & 0.14 & $+$0.08 & 0.14 \\
\ion{Ni}{i}  &    28 &    3.65 & $-$2.57 & 0.11 & $+$0.02 & 0.11 \\
\ion{Ni}{ii} &     4 &    3.90 & $-$2.32 & 0.31 & $+$0.27 & 0.31 \\
\ion{Cu}{i}  &     1 & $<$0.91 &$<-$3.28 &\dots &$<-$0.69 &\dots \\
\ion{Zn}{ii} &     3 &    2.33 & $-$2.23 & 0.11 & $+$0.36 & 0.11 \\
\ion{Sr}{ii} &     3 &    0.60 & $-$2.27 & 0.12 & $+$0.32 & 0.12 \\
\ion{Y}{ii}  &    16 & $-$0.02 & $-$2.23 & 0.11 & $+$0.36 & 0.11 \\
\ion{Zr}{ii} &    26 &    0.68 & $-$1.90 & 0.11 & $+$0.69 & 0.11 \\
\ion{Nb}{ii} &     3 & $-$0.34 & $-$1.80 & 0.18 & $+$0.79 & 0.18 \\
\ion{Mo}{ii} &     4 &    0.14 & $-$1.74 & 0.13 & $+$0.86 & 0.13 \\
\ion{Ru}{i}  &     1 & $<$1.05 &$<-$0.70 &\dots &$<+$1.89 &\dots \\
\ion{Rh}{i}  &     1 & $<$0.04 &$<-$0.87 &\dots &$<+$1.72 &\dots \\
\ion{Pd}{i}  &     1 & $<$0.73 &$<-$0.83 &\dots &$<+$1.76 &\dots \\
\ion{Ag}{i}  &     1 &$<-$0.24 &$<-$1.17 &\dots &$<+$1.42 &\dots \\
\ion{Cd}{i}  &     1 & $-$0.36 & $-$2.07 & 0.26 & $+$0.52 & 0.26 \\
\ion{Te}{i}  &     1 & $<$0.71 &$<-$1.47 &\dots &$<+$1.12 &\dots \\
\ion{Ba}{ii} &     4 &    0.33 & $-$1.85 & 0.16 & $+$0.74 & 0.15 \\
\ion{La}{ii} &    14 & $-$0.58 & $-$1.68 & 0.11 & $+$0.91 & 0.11 \\
\ion{Ce}{ii} &    12 & $-$0.34 & $-$1.92 & 0.10 & $+$0.68 & 0.11 \\
\ion{Pr}{ii} &     3 & $-$0.83 & $-$1.55 & 0.15 & $+$1.05 & 0.16 \\
\ion{Nd}{ii} &    19 & $-$0.31 & $-$1.73 & 0.11 & $+$0.87 & 0.11 \\
\ion{Sm}{ii} &    21 & $-$0.56 & $-$1.52 & 0.11 & $+$1.08 & 0.11 \\
\ion{Eu}{ii} &     7 & $-$0.78 & $-$1.30 & 0.12 & $+$1.29 & 0.12 \\
\ion{Gd}{ii} &    11 & $-$0.26 & $-$1.33 & 0.11 & $+$1.26 & 0.11 \\
\ion{Tb}{ii} &     2 & $-$1.02 & $-$1.32 & 0.25 & $+$1.27 & 0.26 \\
\ion{Dy}{ii} &    13 & $-$0.15 & $-$1.25 & 0.11 & $+$1.34 & 0.11 \\
\ion{Ho}{ii} &     4 & $-$0.70 & $-$1.18 & 0.12 & $+$1.42 & 0.12 \\
\ion{Er}{ii} &     5 & $-$0.46 & $-$1.38 & 0.13 & $+$1.21 & 0.13 \\
\ion{Tm}{ii} &     2 & $-$1.07 & $-$1.17 & 0.23 & $+$1.42 & 0.24 \\
\ion{Yb}{ii} &     1 & $-$0.69 & $-$1.53 & 0.27 & $+$1.06 & 0.26 \\
\ion{Lu}{ii} &     1 & $-$1.16 & $-$1.26 & 0.25 & $+$1.33 & 0.26 \\
\ion{Hf}{ii} &     2 & $-$0.82 & $-$1.67 & 0.25 & $+$0.93 & 0.25 \\
\ion{Os}{ii} &     1 & $-$0.10 & $-$1.50 & 0.27 & $+$1.09 & 0.27 \\
\ion{Ir}{i}  &     1 & $<$0.53 &$<-$0.85 &\dots &$<+$1.74 &\dots \\
\ion{Pt}{i}  &     4 &    0.22 & $-$1.40 & 0.19 & $+$1.20 & 0.19 \\
\ion{Au}{i}  &     2 & $-$0.78 & $-$1.70 & 0.26 & $+$0.89 & 0.26 \\
\ion{Pb}{i}  &     1 & $<$0.83 &$<-$0.92 &\dots &$<+$1.67 &\dots \\
\ion{Th}{ii} &     1 & $-$1.16 & $-$1.18 & 0.23 & $+$1.41 & 0.23 \\
\ion{U}{ii}  &     1 &$<-$0.61 & $<$0.00 &\dots &$<+$2.52 &\dots \\
\\
    \end{tabular}
    }
\tablefoot{All abundances listed in this table are 1D and LTE.}
\end{table}

\begin{table}
\caption{UV abundance summary for J0538.}
\label{tab:abunUV}
\resizebox{\columnwidth}{!}{
    \centering
    \begin{tabular}{lrrrrrr}
    \hline\hline
 El. & N & $\log\epsilon$ (X) & $\mathrm{[X/H]}$ & $\sigma_{\mathrm{[X/H]}}$ & $\mathrm{[X/Fe]}$ & $\sigma_{\mathrm{[X/Fe]}}$ \\
 &  &  &  & [dex] & & [dex]\\
\hline
\ion{Mg}{i}  &  4  &    5.37 & $-$2.23 & 0.11 & $+$0.37 & 0.11\\  
\ion{Al}{i}  &  3  &    3.24 & $-$3.21 & 0.23 & $-$0.62 & 0.23\\	 
\ion{Al}{ii} &  1  &    3.87 & $-$2.58 & 0.23 & $+$0.02 & 0.23\\	 
\ion{Si}{i}  &  3  &    5.21 & $-$2.30 & 0.15 & $+$0.30 & 0.15\\	 
\ion{Si}{ii} &  2  &    5.67 & $-$1.84 & 0.25 & $+$0.75 & 0.25\\	 
\ion{P}{i}   &  1  &    2.69 & $-$2.72 & 0.26 & $-$0.13 & 0.26\\	 
\ion{Ca}{i}  &  1  &    3.81 & $-$2.53 & 0.24 & $+$0.06 & 0.24\\	 
\ion{Sc}{ii} &  1  &    1.20 & $-$1.95 & 0.26 & $+$0.65 & 0.27\\	 
\ion{Ti}{i}  &  3  &    2.75 & $-$2.20 & 0.13 & $+$0.39 & 0.13\\	 
\ion{Ti}{ii} & 13  &    2.86 & $-$2.09 & 0.11 & $+$0.50 & 0.11\\	 
\ion{V}{ii}  & 14  &    1.61 & $-$2.32 & 0.11 & $+$0.27 & 0.11\\	 
\ion{Cr}{i}  &  4  &    2.83 & $-$2.81 & 0.12 & $-$0.21 & 0.12\\	 
\ion{Cr}{ii} & 10  &    3.29 & $-$2.35 & 0.11 & $+$0.24 & 0.12\\	 
\ion{Fe}{i}  & 19  &    4.91 & $-$2.59 & 0.10 & \dots   &\dots\\	 
\ion{Fe}{ii} &  4  &    4.98 & $-$2.52 & 0.11 & $+$0.08 &\dots\\	 
\ion{Co}{ii} &  6  &    2.48 & $-$2.51 & 0.14 & $+$0.08 & 0.14\\	 
\ion{Ni}{i}  & 16  &    3.64 & $-$2.58 & 0.11 & $+$0.01 & 0.11\\
\ion{Ni}{ii} &  4  &    3.90 & $-$2.32 & 0.31 & $+$0.27 & 0.31\\	 
\ion{Y}{ii}  &  1  & $-$0.14 & $-$2.35 & 0.27 & $+$0.24 & 0.27\\	 
\ion{Zr}{ii} & 10  &    0.58 & $-$2.00 & 0.11 & $+$0.59 & 0.12\\	 
\ion{Nb}{ii} &  3  & $-$0.34 & $-$1.80 & 0.18 & $+$0.79 & 0.18\\	 
\ion{Mo}{ii} &  4  &    0.14 & $-$1.74 & 0.13 & $+$0.86 & 0.13\\	 
\ion{Cd}{i}  &  1  & $-$0.36 & $-$2.07 & 0.26 & $+$0.52 & 0.26\\	 
\ion{Gd}{ii} &  5  & $-$0.33 & $-$1.40 & 0.12 & $+$1.19 & 0.13\\	 
\ion{Er}{ii} &  1  & $-$0.41 & $-$1.33 & 0.31 & $+$1.26 & 0.32\\	 
\ion{Lu}{ii} &  1  & $-$1.16 & $-$1.26 & 0.25 & $+$1.33 & 0.26\\	 
\ion{Hf}{ii} &  1  & $-$1.17 & $-$2.02 & 0.83 & $+$0.57 & 0.83\\	 
\ion{Os}{ii} &  1  & $-$0.10 & $-$1.50 & 0.27 & $+$1.09 & 0.27\\	 
\ion{Pt}{i}  &  4  &    0.22 & $-$1.40 & 0.19 & $+$1.20 & 0.19\\	 
\ion{Au}{i}  &  2  & $-$0.78 & $-$1.70 & 0.26 & $+$0.89 & 0.26\\\\
    \end{tabular}
    }
\tablefoot{All abundances listed in this table are 1D and LTE.}
\end{table}

\begin{table}
\caption{Abundance uncertainties for J0538.}
\label{tab:err}
    \centering
    \resizebox{\columnwidth}{!}{%
    \begin{tabular}{lrrrrrr}
    \hline\hline
    El & $\Delta_{T_{\rm eff}}$ & $\Delta_{\log g}$ & $\Delta_{\xi}$ & $\Delta_{\mathrm{[M/H]}}$ & $s_X$\\ 
    & ($\pm$57~K) & $(\pm$0.08~dex) & ($\pm$0.05~kms$^{-1}$) & ($\pm$0.07~dex)\\ 
    & [dex] & [dex] & [dex] & [dex] & [dex]\\
\hline
\ion{O}{i}   &  $-$0.052 & $+$0.031 & $-$0.003 & $-$0.011  & $+$0.100 \\
\ion{Na}{i}  &  $+$0.037 & $+$0.003 & $-$0.067 & $-$0.016  & $+$0.238 \\
\ion{Mg}{i}  &  $+$0.032 & $-$0.011 & $-$0.022 & $+$0.001  & $+$0.139 \\
\ion{Al}{i}  &  $+$0.071 & $-$0.000 & $-$0.015 & $+$0.010  & $+$0.347 \\
\ion{Al}{ii} &  $+$0.025 & $+$0.019 & $+$0.014 & $+$0.011  & $+$0.224 \\
\ion{Si}{i}  &  $+$0.057 & $-$0.012 & $-$0.044 & $-$0.007  & $+$0.156 \\
\ion{Si}{ii} &  $-$0.022 & $-$0.030 & $-$0.061 & $-$0.049  & $+$0.224 \\
\ion{P}{i}   &  $+$0.117 & $-$0.008 & $-$0.008 & $+$0.014  & $+$0.224 \\
\ion{K}{i}   &  $+$0.038 & $-$0.002 & $-$0.004 & $+$0.004  & $+$0.234 \\
\ion{Ca}{i}  &  $+$0.036 & $-$0.002 & $-$0.004 & $+$0.005  & $+$0.125 \\
\ion{Sc}{ii} &  $+$0.013 & $+$0.018 & $-$0.009 & $+$0.002  & $+$0.168 \\
\ion{Ti}{i}  &  $+$0.059 & $-$0.005 & $+$0.001 & $+$0.009  & $+$0.124 \\
\ion{Ti}{ii} &  $+$0.021 & $+$0.021 & $-$0.012 & $+$0.000  & $+$0.142 \\
\ion{V}{ii}  &  $+$0.013 & $+$0.016 & $+$0.001 & $+$0.003  & $+$0.161 \\
\ion{Cr}{i}  &  $+$0.053 & $-$0.007 & $-$0.006 & $+$0.007  & $+$0.143 \\
\ion{Cr}{ii} &  $+$0.005 & $+$0.010 & $-$0.006 & $+$0.000  & $+$0.156 \\
\ion{Mn}{i}  &  $+$0.046 & $-$0.011 & $+$0.001 & $+$0.011  & $+$0.175 \\
\ion{Mn}{ii} &  $+$0.011 & $+$0.010 & $-$0.008 & $+$0.005  & $+$0.142 \\
\ion{Fe}{i}  &  $+$0.049 & $-$0.004 & $-$0.005 & $+$0.005  & $+$0.123 \\
\ion{Fe}{ii} &  $+$0.015 & $+$0.022 & $-$0.013 & $+$0.002  & $+$0.143 \\
\ion{Co}{i}  &  $+$0.068 & $-$0.012 & $+$0.004 & $+$0.010  & $+$0.166 \\
\ion{Co}{ii} &  $+$0.030 & $+$0.043 & $-$0.001 & $+$0.019  & $+$0.201 \\
\ion{Ni}{i}  &  $+$0.049 & $-$0.009 & $-$0.030 & $+$0.000  & $+$0.139 \\
\ion{Ni}{ii} &  $-$0.001 & $+$0.017 & $-$0.047 & $-$0.017  & $+$0.566 \\
\ion{Zn}{ii} &  $+$0.041 & $+$0.000 & $-$0.001 & $+$0.006  & $+$0.100 \\
\ion{Sr}{ii} &  $+$0.015 & $+$0.015 & $+$0.011 & $+$0.014  & $+$0.100 \\
\ion{Y}{ii}  &  $+$0.022 & $+$0.017 & $-$0.002 & $+$0.005  & $+$0.121 \\
\ion{Zr}{ii} &  $+$0.026 & $+$0.017 & $+$0.002 & $+$0.009  & $+$0.150 \\
\ion{Nb}{ii} &  $+$0.048 & $+$0.034 & $+$0.002 & $+$0.017  & $+$0.141 \\
\ion{Mo}{ii} &  $+$0.056 & $+$0.011 & $+$0.002 & $+$0.014  & $+$0.100 \\
\ion{Cd}{i}  &  $+$0.062 & $-$0.006 & $+$0.014 & $+$0.013  & $+$0.224 \\
\ion{Ba}{ii} &  $+$0.038 & $+$0.023 & $-$0.035 & $-$0.005  & $+$0.185 \\
\ion{La}{ii} &  $+$0.023 & $+$0.023 & $+$0.007 & $+$0.011  & $+$0.144 \\
\ion{Ce}{ii} &  $+$0.013 & $+$0.018 & $+$0.021 & $+$0.013  & $+$0.100 \\
\ion{Pr}{ii} &  $+$0.023 & $+$0.029 & $+$0.028 & $+$0.025  & $+$0.141 \\
\ion{Nd}{ii} &  $+$0.015 & $+$0.012 & $+$0.013 & $+$0.012  & $+$0.157 \\
\ion{Sm}{ii} &  $+$0.013 & $+$0.017 & $+$0.010 & $+$0.012  & $+$0.141 \\
\ion{Eu}{ii} &  $+$0.030 & $+$0.020 & $+$0.007 & $+$0.011  & $+$0.143 \\
\ion{Gd}{ii} &  $+$0.003 & $+$0.018 & $+$0.021 & $+$0.010  & $+$0.115 \\
\ion{Tb}{ii} &  $+$0.013 & $+$0.024 & $+$0.029 & $+$0.010  & $+$0.245 \\
\ion{Dy}{ii} &  $+$0.021 & $+$0.020 & $+$0.001 & $+$0.008  & $+$0.134 \\
\ion{Ho}{ii} &  $+$0.035 & $+$0.013 & $+$0.004 & $+$0.009  & $+$0.147 \\
\ion{Er}{ii} &  $+$0.036 & $+$0.014 & $+$0.005 & $+$0.014  & $+$0.153 \\
\ion{Tm}{ii} &  $+$0.000 & $+$0.017 & $+$0.024 & $+$0.003  & $+$0.224 \\
\ion{Yb}{ii} &  $+$0.078 & $+$0.021 & $-$0.016 & $+$0.010  & $+$0.245 \\
\ion{Lu}{ii} &  $-$0.003 & $+$0.004 & $-$0.003 & $+$0.003  & $+$0.245 \\
\ion{Hf}{ii} &  $+$0.002 & $+$0.008 & $+$0.007 & $+$0.008  & $+$0.224 \\
\ion{Os}{ii} &  $+$0.011 & $+$0.016 & $+$0.015 & $+$0.004  & $+$0.224 \\
\ion{Pt}{i}  &  $+$0.061 & $-$0.013 & $+$0.006 & $+$0.010  & $+$0.310 \\
\ion{Au}{i}  &  $+$0.099 & $-$0.002 & $+$0.006 & $+$0.015  & $+$0.245 \\
\ion{Th}{ii} &  $+$0.008 & $+$0.012 & $-$0.000 & $+$0.002  & $+$0.224 \\
							   \\
    \end{tabular}}
\end{table}

 \begin{figure*}[hbt!]
\centering
\includegraphics[width=0.3\linewidth]{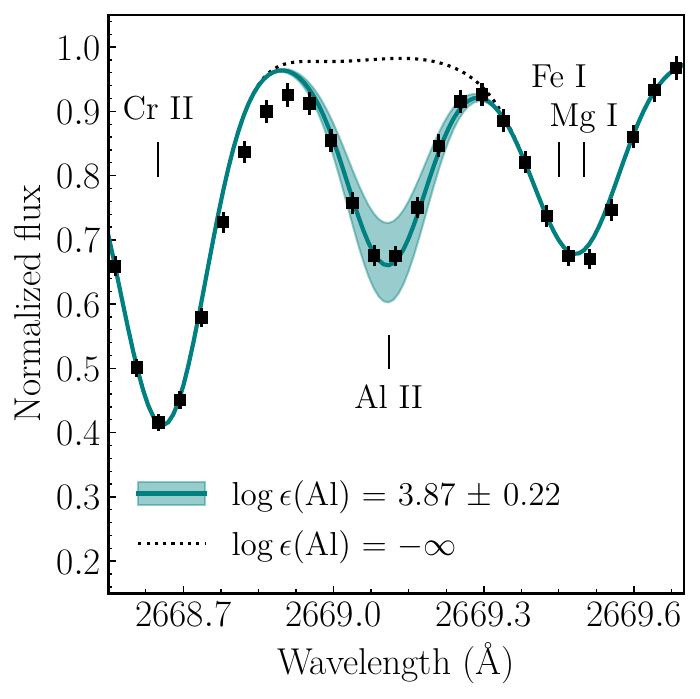}
\includegraphics[width=0.3\linewidth]{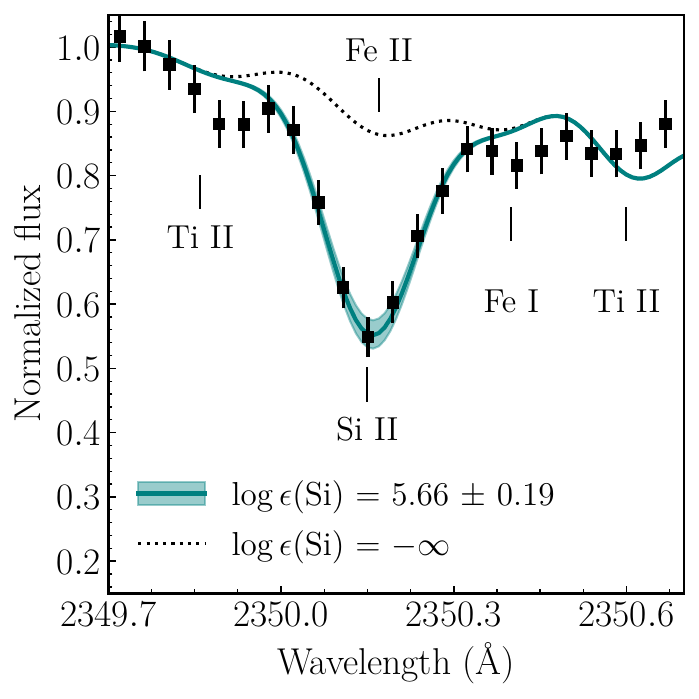}
\includegraphics[width=0.3\linewidth]{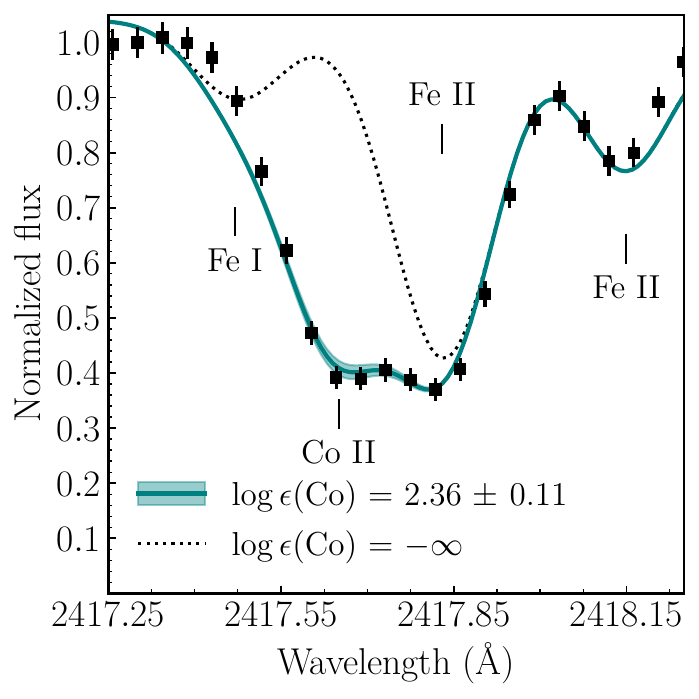}\\
\includegraphics[width=0.3\linewidth]{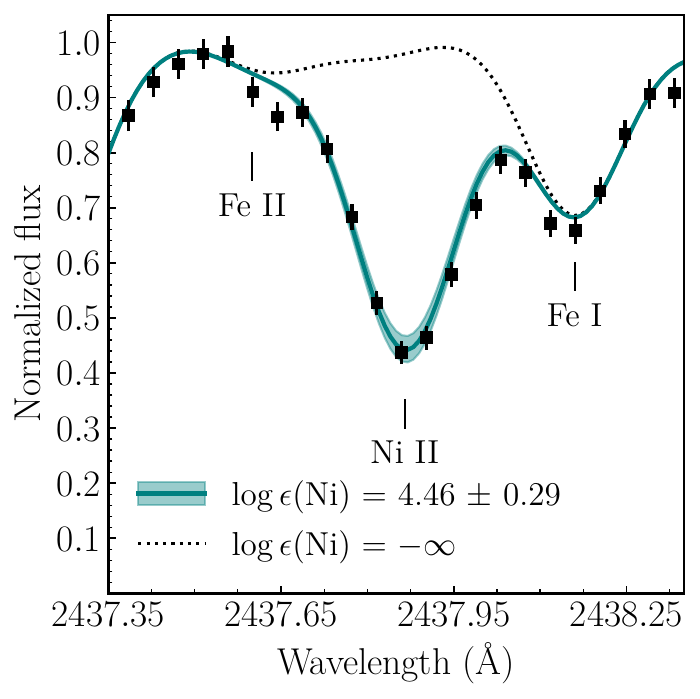}
\includegraphics[width=0.3\linewidth]{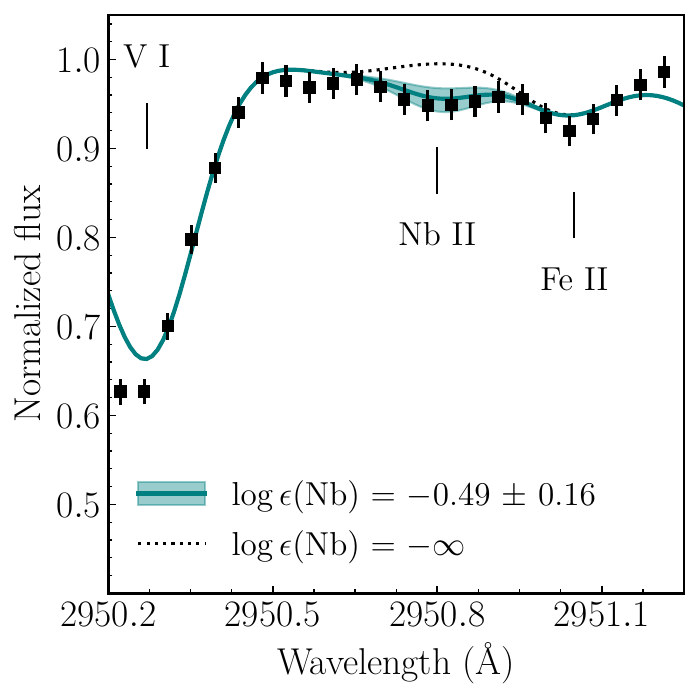}
\includegraphics[width=0.3\linewidth]{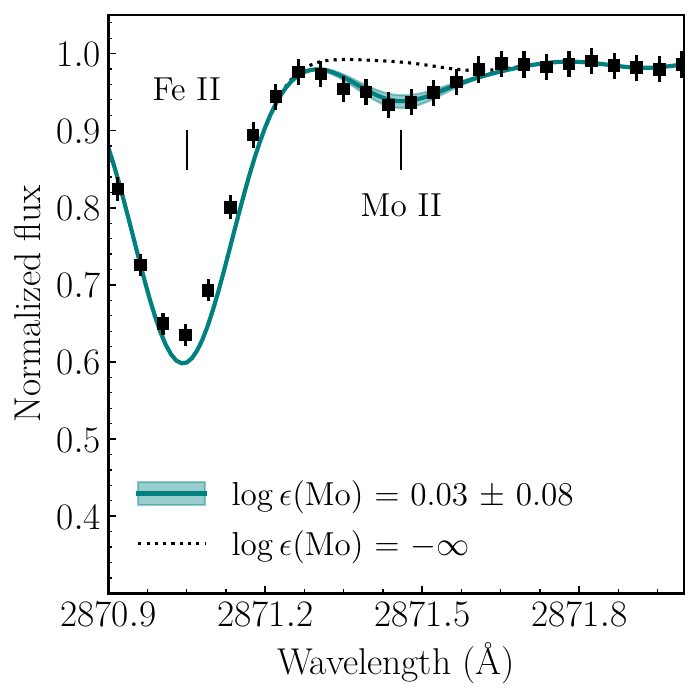}\\
\includegraphics[width=0.3\linewidth]{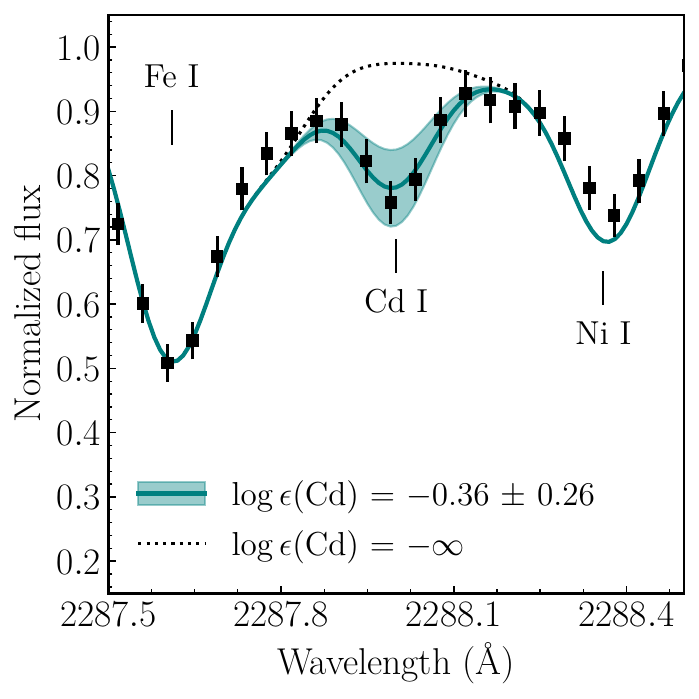}
\includegraphics[width=0.3\linewidth]{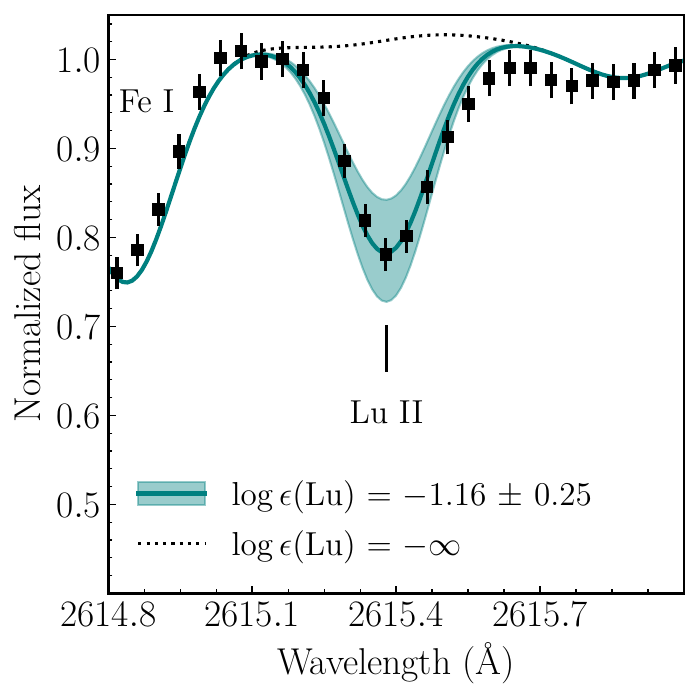}
\includegraphics[width=0.3\linewidth]{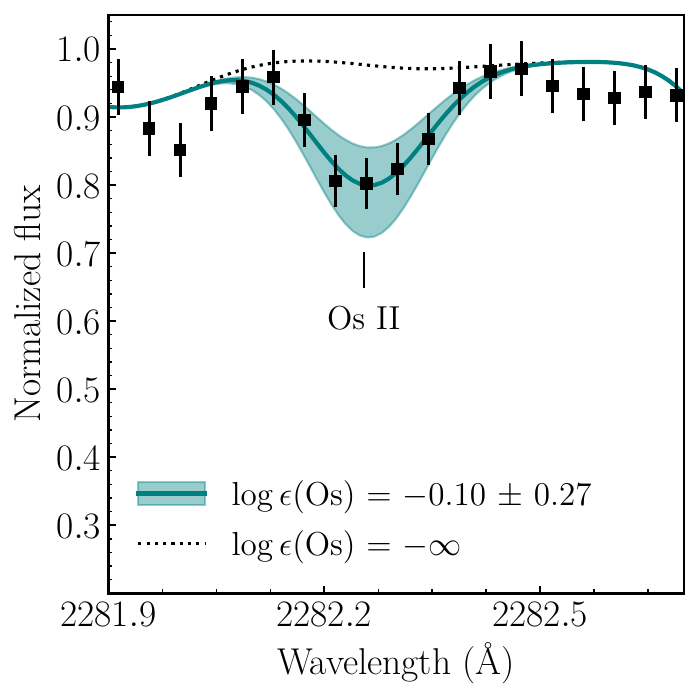}\\
\includegraphics[width=0.3\linewidth]{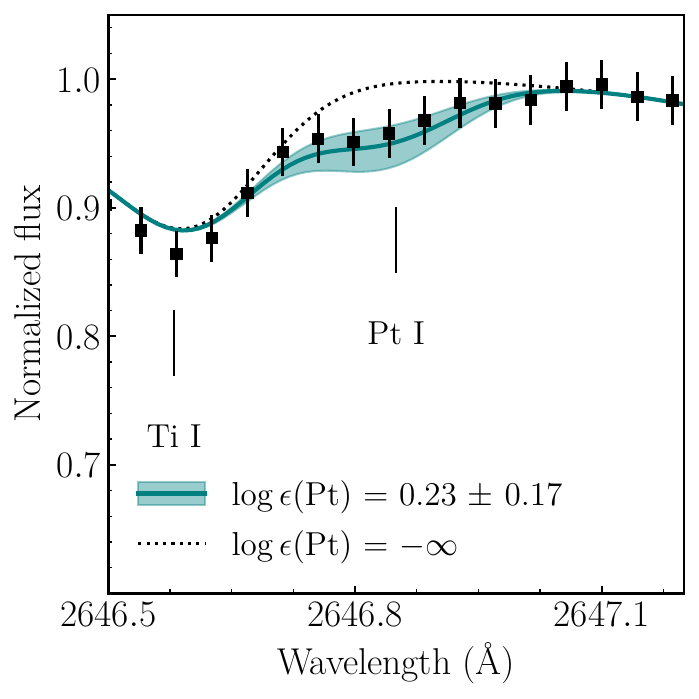}
\includegraphics[width=0.3\linewidth]{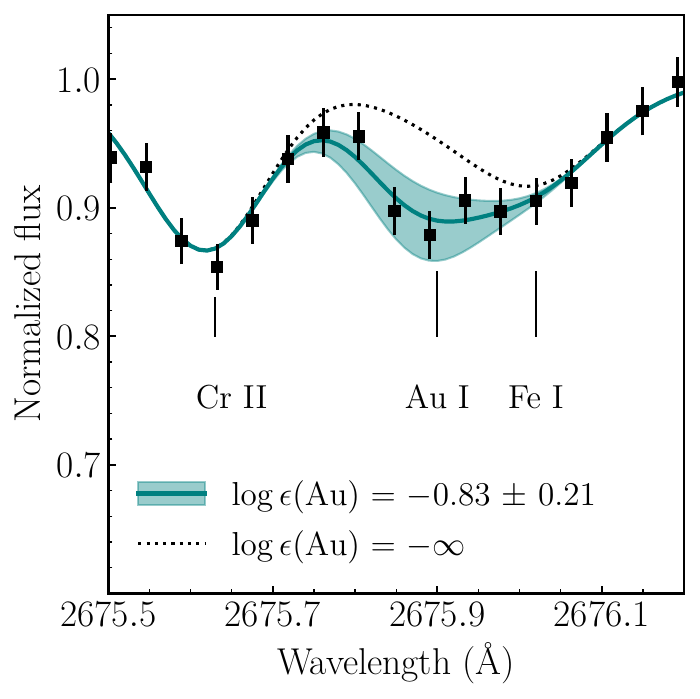}
\includegraphics[width=0.3\linewidth]{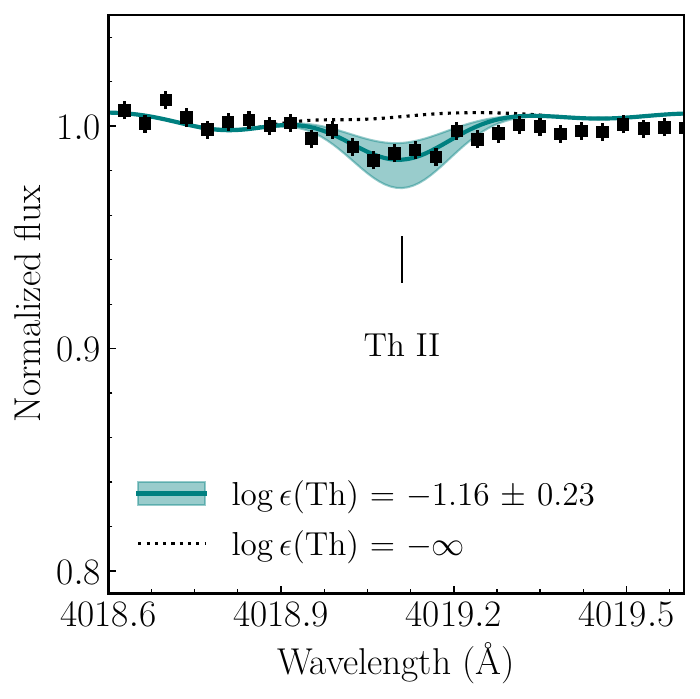}
\caption{Comparison of synthesized and observed spectra (black squares) for selected absorption features in the near-UV spectrum and Th. The blue line is the best-fit synthesis, the blue band shows the uncertainty, and the dotted line is a synthesis without the given element.\label{fig:synth}}
\end{figure*}

\subsection{Elements from oxygen to zinc}
We derive abundances for 17 elements from O to Zn. In addition, we derive 3-$\sigma$ upper limits on C from the CH-$G$ band ($\mathrm{[C/Fe]} < -0.97$), on N from the NH 3360\Angstrom\ band ($\mathrm{[N/Fe]} < +0.77$), and Cu from the 5105\Angstrom\ line ($\mathrm{[Cu/Fe]} < -0.69$). Figure \ref{fig:abun_tot} shows the derived abundances for light elements in J0538 (pink stars) compared to those of the other two RPA stars with HST spectra (HD~222925: green crosses\citep{roederer2022a}, 2MASS~J00512646--1053170: blue squares)\citep{shah2024}, and abundances from MW halo stars (grey dots) \citep{roederer2014a}. For Na and K, we plot the non-LTE corrected abundances. We have corrected the Na and K abundances for J0538 following \citet{lind2011} (Na) and \citet{reggiani2019} (K), resulting in corrections of $-0.58$~dex for Na and $-0.20$~dex for K. The abundances of the light elements in the three RPA HST stars agree quite well with those of the general MW halo population.

For eight elements, we derive abundance estimates from lines of neutral and singly ionized atoms, facilitated by the HST spectrum for the elements Al, Si, Co, and Ni, as shown in Figure \ref{fig:synth}. The abundances derived from the different ionization states deviate for some of these elements, likely due to departures from the LTE assumption employed in this work. For Al, we find $\mathrm{[\ion{Al}{i}/Fe]} = -0.57$ and $\mathrm{[\ion{Al}{ii}/Fe]} = +0.02$, where the ratios indicate the total Al abundance of the star inferred from the neutral and singly ionized species. The difference between our Al abundances corresponds to the non-LTE correction for \ion{Al}{i} of +0.53\,dex found by \citet{lind2022} for the benchmark star with similar parameters, HD~140283 (\teff= 5792\,K, \logg=3.65, $\mathrm{[Fe/H]} = -2.38$). 

\begin{figure*}[hbt!]
\centering
\includegraphics[width=\linewidth]{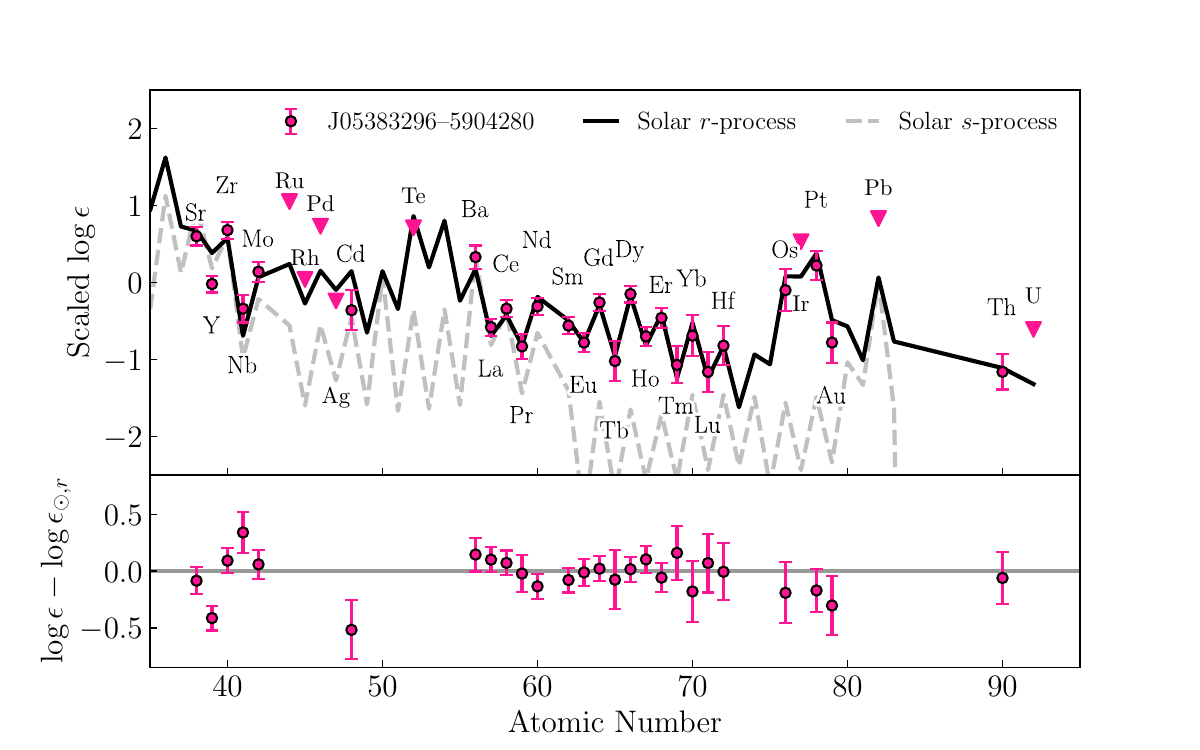}
\caption{Absolute abundances for neutron-capture elements derived for J0538 compared to scaled Solar System \emph{r}- (solid) and $s$-process (dashed) abundances, taken from \citet{sneden2008}. Downwards pointing triangles denote upper limits. The stellar abundances most closely match the Solar System \emph{r}-process abundance pattern.\label{fig:rprofile}}
\end{figure*}

We find a 0.45\,dex difference between our \ion{Si}{i} and \ion{Si}{ii} abundances. The only non-LTE calculations we could find for Si are for the optical 3905\Angstrom\ \ion{Si}{i} line, where the Max Planck Institute for Astronomy (MPIA) based non-LTE correction tool\footnote{\url{https://nlte.mpia.de/gui-siuAC_secE.php}} finds no correction for J0538, based on the grid from \citet{bergemann2013}. It should be noted that \cite{denhartog2023} did not find such large differences in the abundances derived from near-UV \ion{Si}{i} and \ion{Si}{ii} lines for their five metal-poor stars, so the origin of the discrepancy we find is not readily apparent. We find good agreement between our \ion{Ti}{i} and \ion{Ti}{ii} and \ion{Co}{i} and \ion{Co}{ii} abundances, potentially with slightly higher abundances derived from the \ion{Co}{i} lines, similar to the result found be \citet{cowan2020}. The recent work by \cite{mallinson2022}, however, suggests that a non-LTE correction of +0.22\,dex should be applied to the \ion{Ti}{i}, and +0.03\,dex to the \ion{Ti}{ii} abundances of the benchmark star HD~140283 (same as the one used above for Al). In addition, non-LTE corrections for Co as high as +0.6 to +0.8\,dex were found for \ion{Co}{i} lines in metal-poor stars by \cite{Bergemann2010Co}. In our analysis, we find larger offsets between abundances from neutral and single ionized species of Cr, Mn, and Ni compared to Ti and Co. Using the MPIA-based non-LTE correction tool, we find non-LTE corrections of +0.65 to +0.88\,dex for the \ion{Cr}{i} lines used in this analysis, based on the grid from \cite{bergemann2010Cr}, which are significantly larger than the 0.31\,dex abundance offset we find between \ion{Cr}{i} and \ion{Cr}{ii}. Using the same tool, we find corrections of +0.4 to +0.6\,dex for the \ion{Mn}{i} lines used, based on the grid from \cite{bergemann2008}, closer to our abundance offset of 0.24\,dex. Again, the offset we find between abundances derived from neutral and singly ionized lines of Cr and Mn agree with the offsets found in the detailed study of iron-peak elements by \citet{cowan2020}. Finally, \cite{eitner2023} recently computed the non-LTE effects for Ni and determined a correction of +0.15\,dex for \ion{Ni}{i} abundances for stars in our atmospheric parameter regime, bringing our \ion{Ni}{i} abundances closer to the \ion{Ni}{ii} value.

\subsection{Neutron-capture elements}
We derive abundances for 26 neutron-capture elements from Sr to Th, including abundances of Nb, Mo, Cd, Lu, Os, Pt, and Au, exclusively derived from the near-UV HST spectrum. Figure \ref{fig:synth} shows the synthesis of selected lines of these elements. We note that of the five \ion{Sr}{ii} lines we detect in the spectrum at 3464, 4077, 4161, 4215, and 4305\Angstrom, abundances derived from the two commonly used resonance lines (4077 and 4215\Angstrom) are $\sim$1~dex higher than the abundances derived from the remaining three lines. We, therefore, choose to exclude these two lines when calculating the final \ion{Sr}{ii} abundances for the star. Unfortunately, it was not possible to derive abundances or constraining upper limits for the neutron-capture elements Ru, Rh, Pd, Ag, Te, Ir, Pb, and U; we, therefore, provide 3$\sigma$ upper limits for these elements in Table \ref{tab:abunAll}. Figure \ref{fig:rprofile} shows the $\log\epsilon$(X) abundances of the neutron-capture elements detected in J0538 as a function of atomic number compared to the Solar \emph{r}- and \emph{s}-process abundances patterns. A good agreement with the \emph{r}-process pattern is seen, confirming the \emph{r}-process origin of the neutron-capture elements in the atmosphere of this star. This is further supported by the low $\mathrm{[Ba/Eu]}$ ratio of $\mathrm{[Ba/Eu]} = -0.56$ found in the star.

\section{Discussion}
\label{sec:discuss}
\begin{figure*}[hbt!]
\centering
\includegraphics[width=\linewidth]{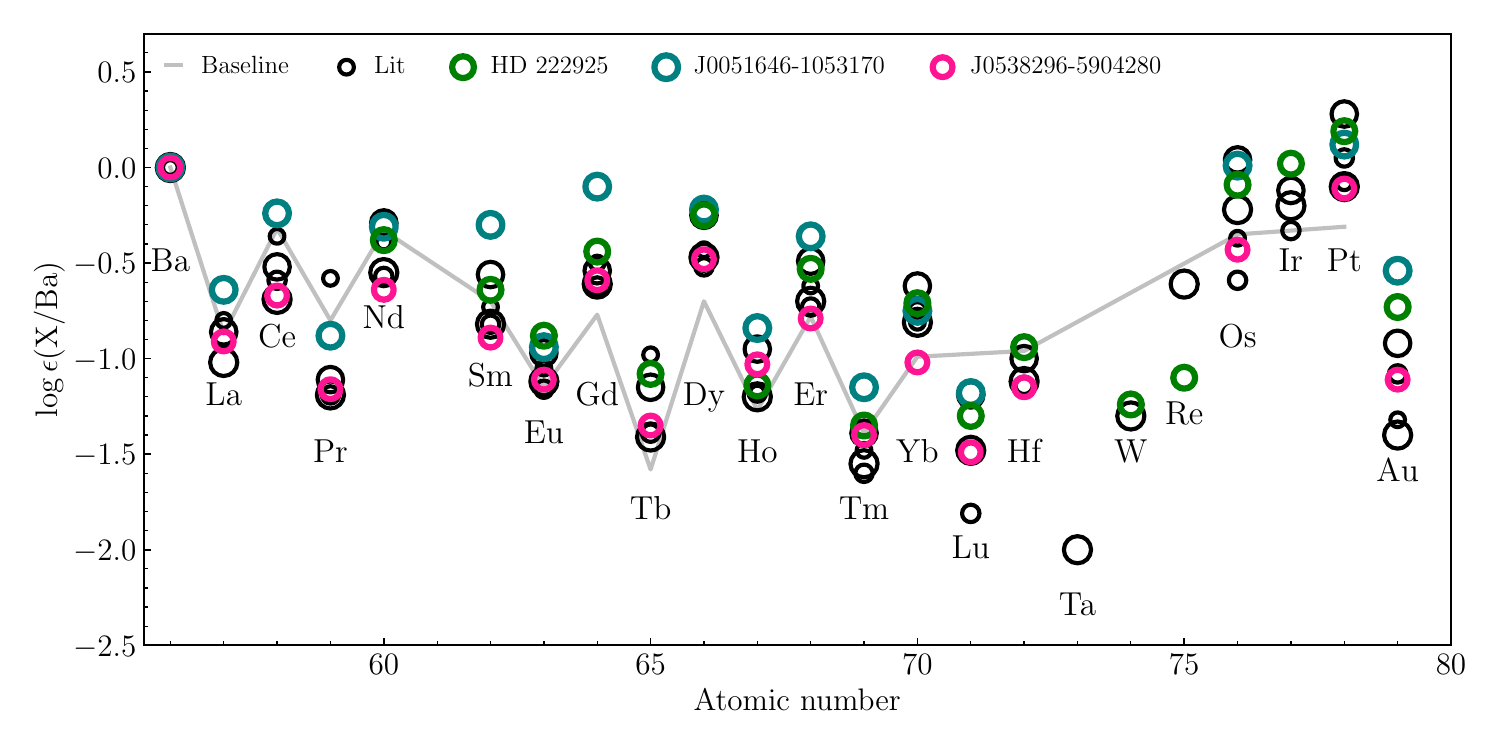}
\caption{Absolute abundances of heavy \emph{r}-process elements in the three RPA stars observed with the HST (pink, green, and blue circles) and literature stars (black circles), with Au detections compared to the baseline pattern from \citet{roederer2023} (grey line), scaled to Ba. The circle size is correlated with the Eu abundance of the star. A somewhat larger star-to-star scatter is seen for Au compared to the other elements. Note that \cite{roederer2023} did not include Lu, Ta, W, Re, and Ir in their sample, hence the baseline pattern (grey line) in Figure \ref{fig:HSTcomb_heavy} is not representative for these elements.  \label{fig:HSTcomb_heavy}}
\end{figure*}

J0538 is the third \emph{r}-II star for which the RPA has obtained near-UV spectra with the HST. In all three stars, we have derived abundances of Cd and Au, increasing the sample of Cd abundances in \emph{r}-process-enriched stars from seven to ten and Au abundances from four to seven. In the following, we discuss the abundance signatures of Cd and Au for \emph{r}-process-enriched stars in relation to two recent RPA discoveries, namely the universal light \emph{r}-process-element pattern \citep{roederer2022b} and the fission-fragment deposition signature \citep{roederer2023}. Figures \ref{fig:HSTcomb_light} and \ref{fig:HSTcomb_heavy} show the neutron-capture element abundances derived for the three RPA HST stars: J0538 (pink circles), HD~222925 (green circles) and 2MASS~J00512646--1053170 (blue circles), along with literature stars (black circles) with Cd (Figure \ref{fig:HSTcomb_light}) and Au (Figure \ref{fig:HSTcomb_heavy}) abundances derived, compared to the ``baseline'' pattern from \cite{roederer2023} (grey lines), determined from \emph{r}-process-enriched stars with $\mathrm{[Eu/Fe]}\leq +0.3$ that do not display the fission fragment enhancement signature. Following \citet{roederer2023}, elements from Sr to Cd are scaled to Zr, while the heavy neutron-capture elements from Ba to Au are scaled to Ba. The circle sizes are correlated with the $\mathrm{[Eu/Fe]}$ ratios, such that the larger circles represent the stars with higher $\mathrm{[Eu/Fe]}$ ratios, and smaller circles represent those with lower $\mathrm{[Eu/Fe]}$ ratios, as in \cite{roederer2023}.

\subsection{Light \emph{r}-process elements, including Cd}
\begin{figure}[hbt!]
\centering
\includegraphics[width=\linewidth]{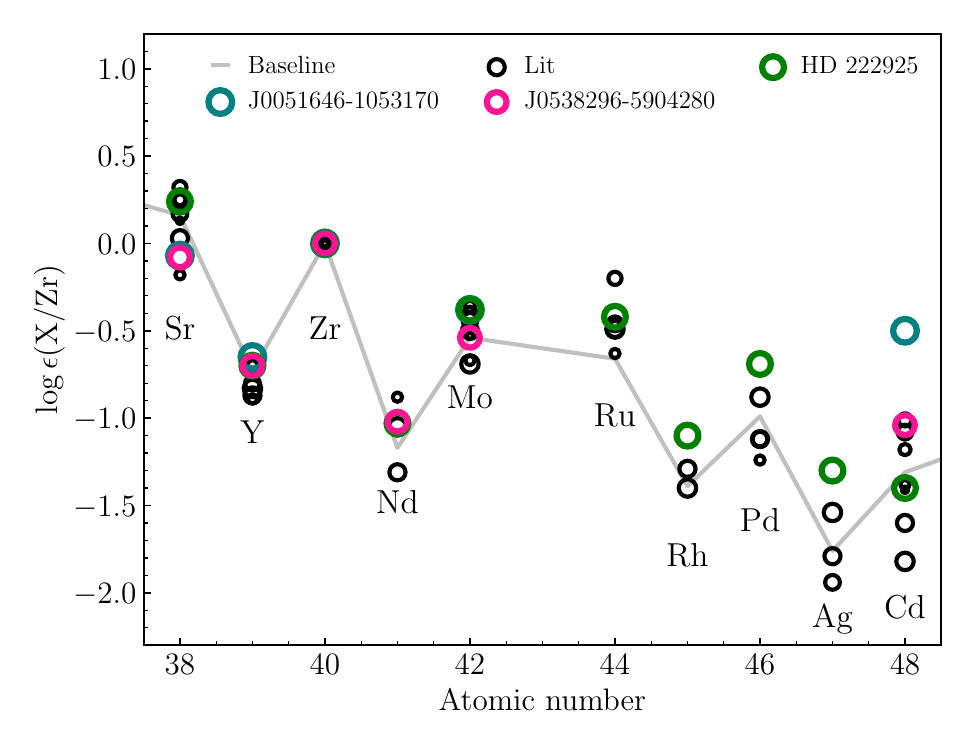}
\caption{Absolute abundances of light \emph{r}-process elements for the three RPA stars observed with the HST (pink, green, and blue circles) and literature stars (black circles) with Cd detections compared to the baseline pattern from \citet{roederer2023} (grey line), scaled to Zr. The circle size is correlated with the Eu abundances of the stars. Elements from Sr to Mo follow the universal light element pattern, while signatures of fission-fragment deposition are seen for Ru to Ag. Cd displays a large scatter and no clear correlation with Eu abundances.  \label{fig:HSTcomb_light}}
\end{figure}

\citet{roederer2022b} showed that, by scaling the light \emph{r}-process elements (Se to Te) to Zr, instead of scaling the full pattern to Eu as in most previous studies \citep[e.g.][]{sneden2008}, a consistent pattern for the elements from Sr to Mo emerges for \emph{r}-process-enriched stars, implying a local universality in the production of these elements. In Figure \ref{fig:HSTcomb_light}, we plot the abundances of the light \emph{r}-process elements in the three RPA HST stars and literature \emph{r}-process-enriched stars with Cd derived abundances. The literature stars included in Figure \ref{fig:HSTcomb_light} are HD~140283 ($\log\epsilon$(Cd) = $-$1.46; \citealt{peterson2020}), HD~128279 ($\log\epsilon$(Cd) = $-$1.38; \citealt{roederer2012,roederer2014b,roederer2022b}), HD~19445 ($\log\epsilon$(Cd) = $-$0.36; \citealt{peterson2020}), HD~84937 ($\log\epsilon$(Cd) = $-$0.36; \citealt{peterson2020}), HD~160617 ($\log\epsilon$(Cd) = $-$0.03; \citealt{roedererlawler2012,peterson2020}), HD~108317 ($\log\epsilon$(Cd) = $-$1.15; \citealt{roederer2012,roederer2014b,roederer2022b}), and BD+17$^\circ$3248 ($\log\epsilon$(Cd) = +0.99; 
\citealt{cowan2002,sneden2009,denhartog2005,roederer2022b}). From the inspection of Figure \ref{fig:HSTcomb_light}, the three RPA HST stars and the literature sample all agree with the recently discovered universality among light \emph{r}-process elements extending up to Mo \citep{roederer2022b}. For the elements from Ru to Ag, Figure \ref{fig:HSTcomb_light} shows a scatter in the abundances, which was attributed by \cite{roederer2023} to fission-fragment deposition. For increasing Eu abundances, the fission-fragment deposition results in an increase of the abundances of these elements above the baseline pattern (grey line), which represents an empirical pattern without fission-fragment deposition. Moving on to the Cd abundances, a much larger spread between the stars and no clear correlation with Eu abundances is seen. This spread is somewhat driven by the very high Cd abundance derived for 2MASS~J00512646--1053170 of $\mathrm{[Cd/Fe]} = +1.27$ by \cite{shah2024}. \cite{shah2024} investigated this spread and found correlations of the derived Cd abundances with the stellar parameters \teff and \logg of these stars, suggesting that non-LTE effects might affect the Cd abundances. The Cd abundance we derive for J0538 of $\mathrm{[Cd/Fe]} = +0.52$, along with the \teff and \logg for J0538, follows the correlations found in \cite{shah2024}, thus supporting the suspected non-LTE effects.

\subsection{Heavy \emph{r}-process elements, including Au}
The behavior of the heavy \emph{r}-process elements is investigated with Figure \ref{fig:HSTcomb_heavy}, where we include the four \emph{r}-process-enriched stars present in the literature with Au abundances preceding RPA work: HD~108317 ($\log\epsilon$(Au) = $-$1.64; \citealt{roederer2012,roederer2014b,roederer2022b}), BD+17$^\circ$3248 ($\log\epsilon$(Au) = $-$0.60; \citealt{roederer2022b}), CS~22892–052 ($\log\epsilon$(Au) = $-$0.90; \citealt{sneden2003}), and CS~31082–001 ($\log\epsilon$(Au) = $-$1.00; \citealt{siqueira2013}). We note that only two literature stars, namely HD~108317 and BD+17$^\circ$3248, appear both in Figures \ref{fig:HSTcomb_light} and \ref{fig:HSTcomb_heavy}; the RPA has thus more than doubled the number of \emph{r}-process-enriched stars for which abundances of both Cd and Au has been derived. Similar to Cd, Figure \ref{fig:HSTcomb_heavy} shows a somewhat larger scatter for Au compared to the other heavy \emph{r}-process elements and no clear correlation between an increased Au abundance and an increased Eu abundance. The standard deviation of the plotted $\log\epsilon$(Au/Ba) ratios is 0.31, similar to the uncertainty on the Au abundances reported by most of the studies included. In comparison, the 1-$\sigma$ values for $\log\epsilon$(Os/Ba), $\log\epsilon$(Ir/Ba), and $\log\epsilon$(Pt/Ba) are 0.24, 0.15, and 0.15 respectively. If we investigate the scatter within the third peak alone, we find 1-$\sigma$ values for $\log\epsilon$(Os/Au) and $\log\epsilon$(Pt/Au) of 0.25 and 0.24, respectively, very close to the value for $\log\epsilon$(Au/Ba), pointing to an element-to-element scatter present within the third-peak elements, and not just with respect to Ba. 

Several studies have found that the shape and position of the third \emph{r}-process peak are highly influenced by the choice of fission model, fission-fragment distribution, and adopted $\beta$-decay rates \citep{surman2001,lund2023}. For example, beta-delayed neutron emission can lead to late-time neutron capture, affecting the third peak \citep{eichler2015}. \cite{roederer2023} found a correlation between the abundance of Eu and the third-peak elements Os and Pt in their sample of \emph{r}-process-enriched stars, suggesting that the ratios of these elements to Eu are possibly affected by fission-fragment deposition or naturally require a more neutron-rich \emph{r}-process for their production in the first place. Au was not included in that study due to the small number of abundances derived for this element. However, the large spread in Au abundances seen in Figure \ref{fig:HSTcomb_heavy} highlights Au as an element needing further investigation.

\begin{figure*}[hbt!]
\centering
\includegraphics[width=0.33\linewidth]{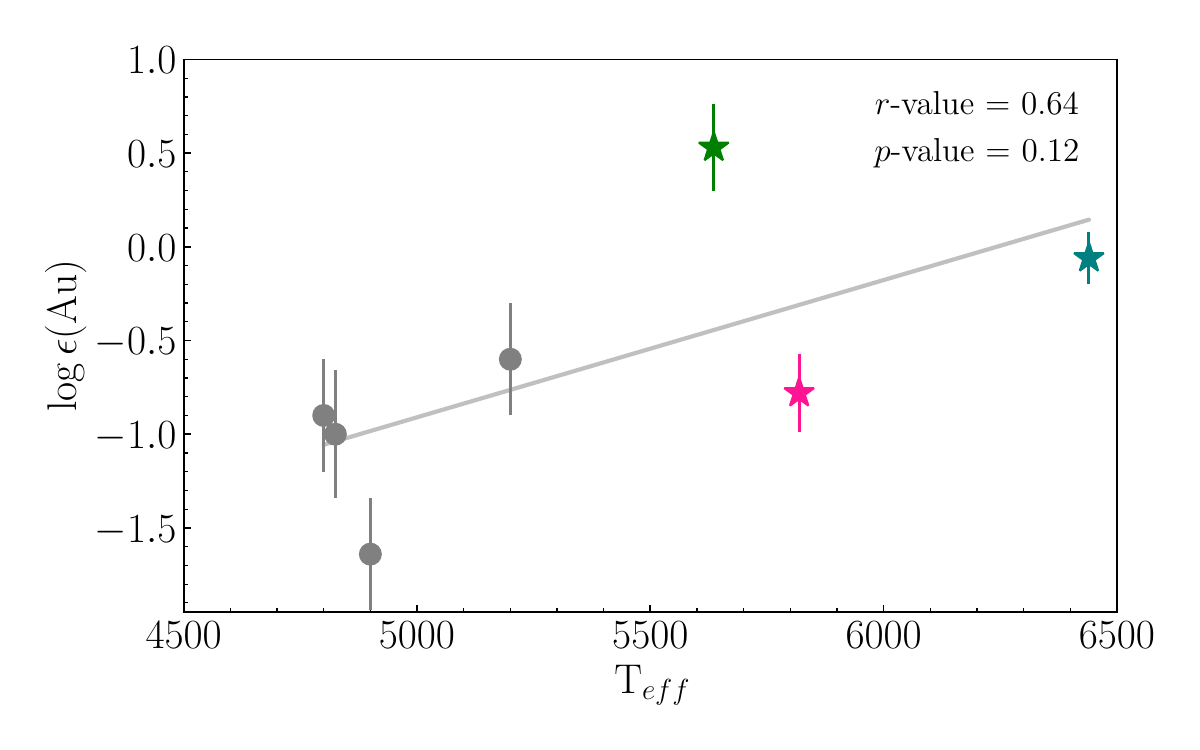}
\includegraphics[width=0.33\linewidth]{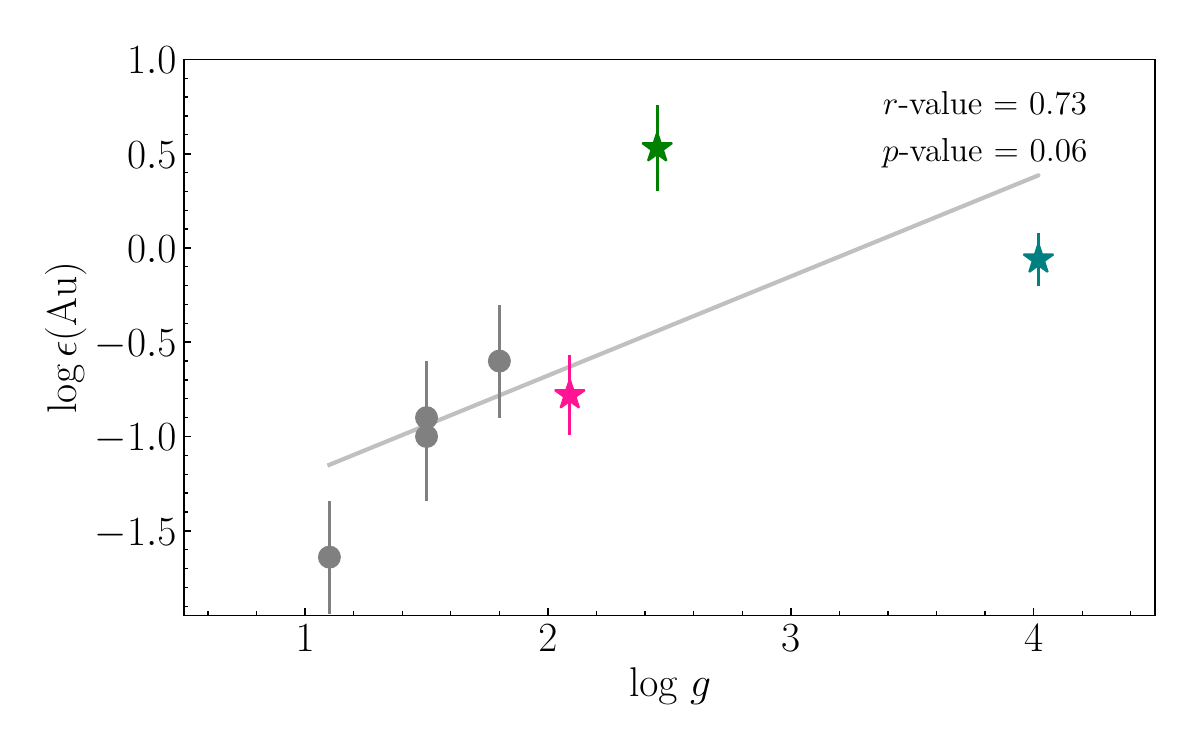}
\includegraphics[width=0.33\linewidth]{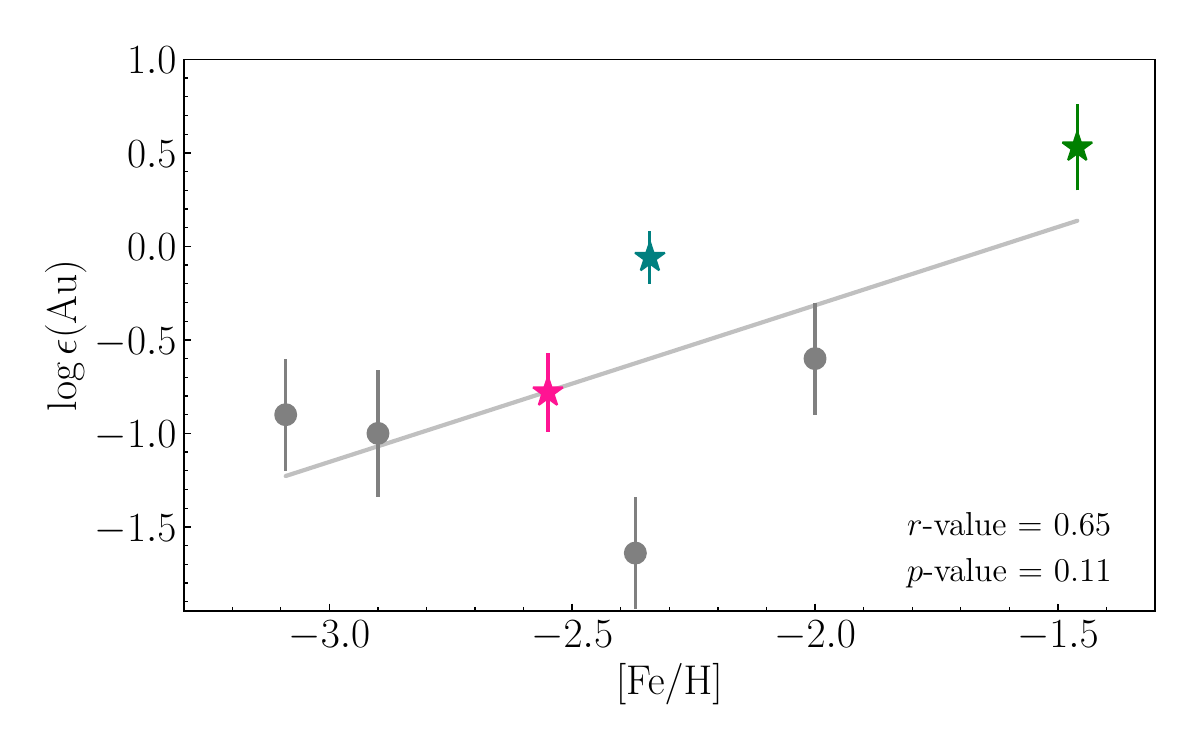}
\caption{Absolute abundances of Au in the three RPA stars (pink, green and blue stars) and literature \emph{r}-process-enriched stars included in Figure \ref{fig:HSTcomb_heavy} (grey dots), as a function of stellar parameters \teff (left panel), \logg (middle panel), and $\mathrm{[Fe/H]}$ (right panel). In each panel, we also plot a least squares fit (grey line) and provide the Pearson correlation coefficient (\emph{r}-value) and the corresponding probability of observing a correlation of at least this magnitude by chance ($p$-value). The abundances of Au seem to correlate with the stellar parameters, hinting at the presence of non-LTE effects. \label{fig:Auparam}}
\end{figure*}

The Au abundances in all of these stars use the \ion{Au}{i} line at 2675\Angstrom, and all but one uses the atomic data from \cite{hannaford1981}, $\log gf = -0.45$; the study of HD~108317 uses atomic data from \cite{fivet2006} with an almost identical $\log gf = -0.47$. However, while this study and that of 2MASS~J00512646--1053170, HD~222925, and BD+17$^\circ$3248 include HFS \citep{demidov2021}, HFS was not considered in the earlier studies of HD~108317, CS~31081-001, and CS~22892-052. We tested the effect of excluding the HFS components from our line list and found a change in the Au abundance of J0538 of $-0.05$\,dex. Hence, this is not likely the source of the spread seen in the Au abundances between these stars. However, the \ion{Au}{i} line at 2675\Angstrom\ is affected by blends of other species, so different choices by authors about how to treat these blends in the spectrum synthesis could affect the final derived abundances. We investigated adding the blending \ion{Fe}{i} lines identified by \citet{ernandes2023} to our 2675 \ion{Au}{i} line list, one of the three \ion{Fe}{i} lines was already included in the list, and found a difference in Au abundance of -0.07~dex for J0538. For completeness, we note that in J0538 and 2MASS~J00512646--1053170, the \ion{Au}{i} line at 2427\Angstrom\ was also used to derive the final Au abundance. In addition, these abundances may also be affected by non-LTE effects. To investigate this, in Figure \ref{fig:Auparam}, we plot the $\log\epsilon$(Au) abundances of the stars included in Figure \ref{fig:HSTcomb_heavy} as a function of the stellar parameters: \teff (left panel), \logg (middle panel), and $\mathrm{[Fe/H]}$ (right panel). For each stellar parameter, we determine the Pearson correlation coefficient ($r$-value) and the corresponding probability of observing a correlation of at least this magnitude by chance ($p$-value), which are shown in the plots. As can be seen in Figure \ref{fig:Auparam}, the $\log\epsilon$(Au) abundances seem to correlate with the stellar parameters for this limited sample of abundances, implying the potential presence of non-LTE effects.

\subsection{Dynamical signature of J0538}
The dynamical properties of J0538 were explored by \cite{gudin2021} and \cite{shank2023} using $Gaia$ data. \cite{gudin2021} derived a pericentric distance of $r_{peri} = 3.23$~kpc, an apocentric distance of $r_{apo} = 9.04$~kpc, and a maximum distance from the Galactic plane of $Z_{max} = 4.81$~kpc for the star using the AGAMA package \citep{vasiliev2019} and the MW2017 potential \citep{mcmillan2017}. Using a friends-of-friends clustering algorithm, \citep{helmi2000}, \citet{gudin2021} also found J0538 to belong to their identified chemo-dynamical tagged group \#6 (CDTG-6). CDTG-6 is a group of three stars that also includes the two \emph{r}-I stars, HD~149414 \citep{fulbright2000} and 2MASS~J01373857--2618131 \citep{holmbeck2020}. Similar kinematic properties for J0538 were found in the analysis from \cite{hattori2023}. However, by employing the greedy optimistic clustering
method \citep{okuno2022}, \cite{hattori2023} pair J0538 in a cluster with 11 other \emph{r}-II stars, which they suggest is a good candidate for a disrupted \emph{r}-process-enhanced dwarf galaxy. In general, kinematic analyses find large numbers of \emph{r}-I and \emph{r}-II stars, including stars from the literature sample used in Figures \ref{fig:HSTcomb_light} and \ref{fig:HSTcomb_heavy}, to belong to dynamical groups characterized by the stars having similar orbital parameters and metallicities \citep{roederer2018b}. Further identification and exploration of these systems will be key to investigating the \emph{r}-process signature in metal-poor stars and further constrain the production of heavy elements in our Universe.

\section{Summary} \label{sec:summary}
This paper presents detailed abundances derived from optical and near-UV spectra of the \emph{r}-II star J0538. This is the third star discovered by the RPA for which near-UV spectra with the HST have been obtained, significantly increasing the sample of \emph{r}-process-enriched stars with abundances derived from near-UV spectra. We derive abundances for 43 elements, including 26 neutron-capture elements, 25 of which are derived from the HST spectrum. In particular, we derive abundances for the neutron-capture elements Cd and Au. Prior to the RPA analysis, only two \emph{r}-process-enriched stars existed in the literature, with abundances reported for both of these elements. We investigate the behaviour of these two elements in \emph{r}-process-enriched stars in relation to the recently discovered fission-fragment deposition signature potentially affecting the neutron-capture element abundances up to the third-peak elements Os and Pt. For both Cd and Au, a larger star-to-star scatter is detected, and no clear correlation with Eu abundances is seen, as is found for other elements affected by fission-fragment deposition. However, \cite{shah2024} found correlations between the derived Cd abundances and the stellar parameters, suggesting that current Cd abundances are affected by non-LTE effects. We find similar correlations between the Au abundances and stellar parameters, suggesting that non-LTE effects also influence the Au abundances similarly derived from neutral species. Therefore, until non-LTE effects for Cd and Au are investigated thoroughly, it will be unclear whether the range of \emph{r}-process elements (Ru to Ag and Gd to Pt) impacted by fission-fragment distribution of transuranic elements extends to elements Cd and Au. This work highlights the importance of the RPA efforts and the need for further near-UV observations of stars with an \emph{r}-process abundance signature in order to more fully understand the formation and evolution of these elements in our Universe.

\begin{acknowledgements}
We thank the referee for a timely report and helpful comments on the paper. 

We acknowledge generous support provided by NASA through grant GO-15951 from the Space Telescope Science Institute, which is operated by the Association of Universities for Research in Astronomy, Incorporated, under NASA contract NAS5-26555.

TTH acknowledges support from the Swedish Research Council (VR 2021-05556) and HST-GO-15657.
TCB acknowledges partial support from grants PHY 14-30152; Physics Frontier Center/JINA Center for the Evolution of the Elements (JINA-CEE), and OISE-1927130: The International Research Network for Nuclear Astrophysics (IReNA), awarded by the US National Science Foundation.  AF acknowledges support from NSF-AAG grant AST-2307436. 
EMH acknowledges work performed under the auspices of the U.S.\ Department of Energy by Lawrence Livermore National Laboratory under Contract DE-AC52-07NA27344.
This document has been approved with release number LLNL-JRNL-2001281-DRAFT.
The work of VMP is supported by NOIRLab, which is managed by the
Association of Universities for Research in Astronomy (AURA) under a
cooperative agreement with the U.S. National Science Foundation.
I.U.R. acknowledges additional funding support from HST-GO-15657 and the 
U.S. National Science Foundation (NSF) grant AST~2205847.
R.E. acknowledges support from NSF-AAG grant AST 2206263 and NASA Astrophysics Theory Program grant 80NSSC24K0899.

This research has made use of NASA’s Astrophysics Data System Bibliographic Services; the  \href{https://arxiv.org/}{arXiv.org} preprint server operated by Cornell University; the SIMBAD and VizieR databases hosted by the Strasbourg Astronomical Data Center \citep{wenger2000}; the ASD hosted by NIST; the MAST at STScI; Image Reduction and Analysis Facility (IRAF) NOIRLab IRAF is distributed by the Community Science and Data Center at NSF NOIRLab, which is managed by the Association of Universities for Research in Astronomy (AURA) under a cooperative agreement with the U.S. National Science Foundation \citep{tody1986,tody1993,fitzpatrick2024}; NumPy \citep{numpy}; Matplotlib \citep{matplotlib}; and AstroPy \citep{Astropy:2013,Astropy:2018}

\end{acknowledgements}

\bibliographystyle{aa}
\bibliography{Terese}

\begin{thebibliography}{159}
\expandafter\ifx\csname natexlab\endcsname\relax\def\natexlab#1{#1}\fi

\bibitem[{{Abbott} {et~al.}(2017){Abbott}, {Abbott}, {Abbott}, {Acernese}, {Ackley}, {Adams}, {Adams}, {Addesso}, {Adhikari}, {Adya}, \& et~al.}]{abbott2017}
{Abbott}, B.~P., {Abbott}, R., {Abbott}, T.~D., {et~al.} 2017, Physical Review Letters, 119, 161101

\bibitem[{{Asplund} {et~al.}(2009){Asplund}, {Grevesse}, {Sauval}, \& {Scott}}]{asplund2009}
{Asplund}, M., {Grevesse}, N., {Sauval}, A.~J., \& {Scott}, P. 2009, \araa, 47, 481

\bibitem[{{Astropy Collaboration} {et~al.}(2018){Astropy Collaboration}, {Price-Whelan}, {Sip{\H{o}}cz}, {G{\"u}nther}, {Lim}, {Crawford}, {Conseil}, {Shupe}, {Craig}, {Dencheva}, {Ginsburg}, {VanderPlas}, {Bradley}, {P{\'e}rez-Su{\'a}rez}, {de Val-Borro}, {Aldcroft}, {Cruz}, {Robitaille}, {Tollerud}, {Ardelean}, {Babej}, {Bach}, {Bachetti}, {Bakanov}, {Bamford}, {Barentsen}, {Barmby}, {Baumbach}, {Berry}, {Biscani}, {Boquien}, {Bostroem}, {Bouma}, {Brammer}, {Bray}, {Breytenbach}, {Buddelmeijer}, {Burke}, {Calderone}, {Cano Rodr{\'\i}guez}, {Cara}, {Cardoso}, {Cheedella}, {Copin}, {Corrales}, {Crichton}, {D'Avella}, {Deil}, {Depagne}, {Dietrich}, {Donath}, {Droettboom}, {Earl}, {Erben}, {Fabbro}, {Ferreira}, {Finethy}, {Fox}, {Garrison}, {Gibbons}, {Goldstein}, {Gommers}, {Greco}, {Greenfield}, {Groener}, {Grollier}, {Hagen}, {Hirst}, {Homeier}, {Horton}, {Hosseinzadeh}, {Hu}, {Hunkeler}, {Ivezi{\'c}}, {Jain}, {Jenness}, {Kanarek}, {Kendrew}, {Kern}, {Kerzendorf}, {Khvalko}, {King}, {Kirkby}, {Kulkarni},
  {Kumar}, {Lee}, {Lenz}, {Littlefair}, {Ma}, {Macleod}, {Mastropietro}, {McCully}, {Montagnac}, {Morris}, {Mueller}, {Mumford}, {Muna}, {Murphy}, {Nelson}, {Nguyen}, {Ninan}, {N{\"o}the}, {Ogaz}, {Oh}, {Parejko}, {Parley}, {Pascual}, {Patil}, {Patil}, {Plunkett}, {Prochaska}, {Rastogi}, {Reddy Janga}, {Sabater}, {Sakurikar}, {Seifert}, {Sherbert}, {Sherwood-Taylor}, {Shih}, {Sick}, {Silbiger}, {Singanamalla}, {Singer}, {Sladen}, {Sooley}, {Sornarajah}, {Streicher}, {Teuben}, {Thomas}, {Tremblay}, {Turner}, {Terr{\'o}n}, {van Kerkwijk}, {de la Vega}, {Watkins}, {Weaver}, {Whitmore}, {Woillez}, {Zabalza}, \& {Astropy Contributors}}]{Astropy:2018}
{Astropy Collaboration}, {Price-Whelan}, A.~M., {Sip{\H{o}}cz}, B.~M., {et~al.} 2018, \aj, 156, 123

\bibitem[{{Astropy Collaboration} {et~al.}(2013){Astropy Collaboration}, {Robitaille}, {Tollerud}, {Greenfield}, {Droettboom}, {Bray}, {Aldcroft}, {Davis}, {Ginsburg}, {Price-Whelan}, {Kerzendorf}, {Conley}, {Crighton}, {Barbary}, {Muna}, {Ferguson}, {Grollier}, {Parikh}, {Nair}, {Unther}, {Deil}, {Woillez}, {Conseil}, {Kramer}, {Turner}, {Singer}, {Fox}, {Weaver}, {Zabalza}, {Edwards}, {Azalee Bostroem}, {Burke}, {Casey}, {Crawford}, {Dencheva}, {Ely}, {Jenness}, {Labrie}, {Lim}, {Pierfederici}, {Pontzen}, {Ptak}, {Refsdal}, {Servillat}, \& {Streicher}}]{Astropy:2013}
{Astropy Collaboration}, {Robitaille}, T.~P., {Tollerud}, E.~J., {et~al.} 2013, \aap, 558, A33

\bibitem[{{Babusiaux} {et~al.}(2023){Babusiaux}, {Fabricius}, {Khanna}, {Muraveva}, {Reyl{\'e}}, {Spoto}, {Vallenari}, {Luri}, {Arenou}, {{\'A}lvarez}, {Anders}, {Antoja}, {Balbinot}, {Barache}, {Bauchet}, {Bossini}, {Busonero}, {Cantat-Gaudin}, {Carrasco}, {Dafonte}, {Diakit{\'e}}, {Figueras}, {Garcia-Gutierrez}, {Garofalo}, {Helmi}, {Jim{\'e}nez-Arranz}, {Jordi}, {Kervella}, {Kostrzewa-Rutkowska}, {Leclerc}, {Licata}, {Manteiga}, {Masip}, {Mongui{\'o}}, {Ramos}, {Robichon}, {Robin}, {Romero-G{\'o}mez}, {S{\'a}ez}, {Santove{\~n}a}, {Spina}, {Torralba Elipe}, \& {Weiler}}]{Gaiared}
{Babusiaux}, C., {Fabricius}, C., {Khanna}, S., {et~al.} 2023, \aap, 674, A32

\bibitem[{{Bailer-Jones} {et~al.}(2021){Bailer-Jones}, {Rybizki}, {Fouesneau}, {Demleitner}, \& {Andrae}}]{bailorjones2021}
{Bailer-Jones}, C.~A.~L., {Rybizki}, J., {Fouesneau}, M., {Demleitner}, M., \& {Andrae}, R. 2021, \aj, 161, 147

\bibitem[{{Bandyopadhyay} {et~al.}(2024){Bandyopadhyay}, {Ezzeddine}, {Allende Prieto}, {Aria}, {Shah}, {Beers}, {Frebel}, {Hansen}, {Holmbeck}, {Placco}, {Roederer}, \& {Sakari}}]{bandyopadhyay2024}
{Bandyopadhyay}, A., {Ezzeddine}, R., {Allende Prieto}, C., {et~al.} 2024, \apjs, 274, 39

\bibitem[{{Barbuy} {et~al.}(2011){Barbuy}, {Spite}, {Hill}, {Primas}, {Plez}, {Cayrel}, {Spite}, {Wanajo}, {Siqueira Mello}, {Andersen}, {Nordstr{\"o}m}, {Beers}, {Bonifacio}, {Fran{\c{c}}ois}, \& {Molaro}}]{barbuy2011}
{Barbuy}, B., {Spite}, M., {Hill}, V., {et~al.} 2011, \aap, 534, A60

\bibitem[{{Belmonte} {et~al.}(2017){Belmonte}, {Pickering}, {Ruffoni}, {Den Hartog}, {Lawler}, {Guzman}, \& {Heiter}}]{belmonte2017}
{Belmonte}, M.~T., {Pickering}, J.~C., {Ruffoni}, M.~P., {et~al.} 2017, \apj, 848, 125

\bibitem[{{Bergemann} \& {Cescutti}(2010)}]{bergemann2010Cr}
{Bergemann}, M. \& {Cescutti}, G. 2010, \aap, 522, A9

\bibitem[{{Bergemann} \& {Gehren}(2008)}]{bergemann2008}
{Bergemann}, M. \& {Gehren}, T. 2008, \aap, 492, 823

\bibitem[{{Bergemann} {et~al.}(2013){Bergemann}, {Kudritzki}, {W{\"u}rl}, {Plez}, {Davies}, \& {Gazak}}]{bergemann2013}
{Bergemann}, M., {Kudritzki}, R.-P., {W{\"u}rl}, M., {et~al.} 2013, \apj, 764, 115

\bibitem[{{Bergemann} {et~al.}(2010){Bergemann}, {Pickering}, \& {Gehren}}]{Bergemann2010Co}
{Bergemann}, M., {Pickering}, J.~C., \& {Gehren}, T. 2010, \mnras, 401, 1334

\bibitem[{{Bernstein} {et~al.}(2003){Bernstein}, {Shectman}, {Gunnels}, {Mochnacki}, \& {Athey}}]{bernstein2003}
{Bernstein}, R., {Shectman}, S.~A., {Gunnels}, S.~M., {Mochnacki}, S., \& {Athey}, A.~E. 2003, in Society of Photo-Optical Instrumentation Engineers (SPIE) Conference Series, Vol. 4841, Instrument Design and Performance for Optical/Infrared Ground-based Telescopes, ed. M.~{Iye} \& A.~F.~M. {Moorwood}, 1694--1704

\bibitem[{{Bi{\'e}mont} {et~al.}(2011){Bi{\'e}mont}, {Blagoev}, {Engstr{\"o}m}, {Hartman}, {Lundberg}, {Malcheva}, {Nilsson}, {Whitehead}, {Palmeri}, \& {Quinet}}]{biemont2011}
{Bi{\'e}mont}, {\'E}., {Blagoev}, K., {Engstr{\"o}m}, L., {et~al.} 2011, \mnras, 414, 3350

\bibitem[{{Bowen} \& {Vaughan}(1973)}]{bowen1973}
{Bowen}, I.~S. \& {Vaughan}, A.~H., J. 1973, \ao, 12, 1430

\bibitem[{{Burbidge} {et~al.}(1957){Burbidge}, {Burbidge}, {Fowler}, \& {Hoyle}}]{burbidge1957}
{Burbidge}, E.~M., {Burbidge}, G.~R., {Fowler}, W.~A., \& {Hoyle}, F. 1957, Reviews of Modern Physics, 29, 547

\bibitem[{{Cain} {et~al.}(2018){Cain}, {Frebel}, {Gull}, {Ji}, {Placco}, {Beers}, {Mel{\'e}ndez}, {Ezzeddine}, {Casey}, {Hansen}, {Roederer}, \& {Sakari}}]{cain2018}
{Cain}, M., {Frebel}, A., {Gull}, M., {et~al.} 2018, \apj, 864, 43

\bibitem[{{Cain} {et~al.}(2020){Cain}, {Frebel}, {Ji}, {Placco}, {Ezzeddine}, {Roederer}, {Hattori}, {Beers}, {Mel{\'e}ndez}, {Hansen}, \& {Sakari}}]{cain2020}
{Cain}, M., {Frebel}, A., {Ji}, A.~P., {et~al.} 2020, \apj, 898, 40

\bibitem[{{Cameron}(1957)}]{cameron1957}
{Cameron}, A.~G.~W. 1957, \pasp, 69, 201

\bibitem[{{Casagrande} \& {VandenBerg}(2014)}]{casagrande2014}
{Casagrande}, L. \& {VandenBerg}, D.~A. 2014, \mnras, 444, 392

\bibitem[{{Castelli} \& {Kurucz}(2003)}]{castelli2003}
{Castelli}, F. \& {Kurucz}, R.~L. 2003, in Modelling of Stellar Atmospheres, ed. N.~{Piskunov}, W.~W. {Weiss}, \& D.~F. {Gray}, Vol. 210, A20

\bibitem[{{C{\^o}t{\'e}} {et~al.}(2017){C{\^o}t{\'e}}, {Belczynski}, {Fryer}, {Ritter}, {Paul}, {Wehmeyer}, \& {O'Shea}}]{cote2017}
{C{\^o}t{\'e}}, B., {Belczynski}, K., {Fryer}, C.~L., {et~al.} 2017, \apj, 836, 230

\bibitem[{{C{\^o}t{\'e}} {et~al.}(2018){C{\^o}t{\'e}}, {Fryer}, {Belczynski}, {Korobkin}, {Chru{\'s}li{\'n}ska}, {Vassh}, {Mumpower}, {Lippuner}, {Sprouse}, {Surman}, \& {Wollaeger}}]{cote2018}
{C{\^o}t{\'e}}, B., {Fryer}, C.~L., {Belczynski}, K., {et~al.} 2018, \apj, 855, 99

\bibitem[{{Cowan} {et~al.}(2002){Cowan}, {Sneden}, {Burles}, {Ivans}, {Beers}, {Truran}, {Lawler}, {Primas}, {Fuller}, {Pfeiffer}, \& {Kratz}}]{cowan2002}
{Cowan}, J.~J., {Sneden}, C., {Burles}, S., {et~al.} 2002, \apj, 572, 861

\bibitem[{{Cowan} {et~al.}(2020){Cowan}, {Sneden}, {Roederer}, {Lawler}, {Hartog}, {Sobeck}, \& {Boesgaard}}]{cowan2020}
{Cowan}, J.~J., {Sneden}, C., {Roederer}, I.~U., {et~al.} 2020, \apj, 890, 119

\bibitem[{{Cutri} {et~al.}(2003){Cutri}, {Skrutskie}, {van Dyk}, {Beichman}, {Carpenter}, {Chester}, {Cambresy}, {Evans}, {Fowler}, {Gizis}, {Howard}, {Huchra}, {Jarrett}, {Kopan}, {Kirkpatrick}, {Light}, {Marsh}, {McCallon}, {Schneider}, {Stiening}, {Sykes}, {Weinberg}, {Wheaton}, {Wheelock}, \& {Zacarias}}]{cutri2003}
{Cutri}, R.~M., {Skrutskie}, M.~F., {van Dyk}, S., {et~al.} 2003, VizieR Online Data Catalog, II/246

\bibitem[{{Demidov} {et~al.}(2021){Demidov}, {Konovalova}, {Imanbaeva}, {Kozlov}, \& {Barzakh}}]{demidov2021}
{Demidov}, Y.~A., {Konovalova}, E.~A., {Imanbaeva}, R.~T., {Kozlov}, M.~G., \& {Barzakh}, A.~E. 2021, \pra, 103, 032824

\bibitem[{{Den Hartog} {et~al.}(2005){Den Hartog}, {Herd}, {Lawler}, {Sneden}, {Cowan}, \& {Beers}}]{denhartog2005}
{Den Hartog}, E.~A., {Herd}, M.~T., {Lawler}, J.~E., {et~al.} 2005, \apj, 619, 639

\bibitem[{{Den Hartog} {et~al.}(2020){Den Hartog}, {Lawler}, \& {Roederer}}]{denhartog2020}
{Den Hartog}, E.~A., {Lawler}, J.~E., \& {Roederer}, I.~U. 2020, \apjs, 248, 10

\bibitem[{{Den Hartog} {et~al.}(2021{\natexlab{a}}){Den Hartog}, {Lawler}, \& {Roederer}}]{denhartog2021a}
{Den Hartog}, E.~A., {Lawler}, J.~E., \& {Roederer}, I.~U. 2021{\natexlab{a}}, \apjs, 254, 5

\bibitem[{{Den Hartog} {et~al.}(2003){Den Hartog}, {Lawler}, {Sneden}, \& {Cowan}}]{denhartog2003}
{Den Hartog}, E.~A., {Lawler}, J.~E., {Sneden}, C., \& {Cowan}, J.~J. 2003, \apjs, 148, 543

\bibitem[{{Den Hartog} {et~al.}(2006){Den Hartog}, {Lawler}, {Sneden}, \& {Cowan}}]{denhartog2006}
{Den Hartog}, E.~A., {Lawler}, J.~E., {Sneden}, C., \& {Cowan}, J.~J. 2006, \apjs, 167, 292

\bibitem[{{Den Hartog} {et~al.}(2019){Den Hartog}, {Lawler}, {Sneden}, {Cowan}, \& {Brukhovesky}}]{denhartog2019}
{Den Hartog}, E.~A., {Lawler}, J.~E., {Sneden}, C., {Cowan}, J.~J., \& {Brukhovesky}, A. 2019, \apjs, 243, 33

\bibitem[{{Den Hartog} {et~al.}(2021{\natexlab{b}}){Den Hartog}, {Lawler}, {Sneden}, {Cowan}, {Roederer}, \& {Sobeck}}]{denhartog2021b}
{Den Hartog}, E.~A., {Lawler}, J.~E., {Sneden}, C., {et~al.} 2021{\natexlab{b}}, \apjs, 255, 27

\bibitem[{{Den Hartog} {et~al.}(2023){Den Hartog}, {Lawler}, {Sneden}, {Roederer}, \& {Cowan}}]{denhartog2023}
{Den Hartog}, E.~A., {Lawler}, J.~E., {Sneden}, C., {Roederer}, I.~U., \& {Cowan}, J.~J. 2023, \apjs, 265, 42

\bibitem[{{Den Hartog} {et~al.}(2011){Den Hartog}, {Lawler}, {Sobeck}, {Sneden}, \& {Cowan}}]{denhartog2011}
{Den Hartog}, E.~A., {Lawler}, J.~E., {Sobeck}, J.~S., {Sneden}, C., \& {Cowan}, J.~J. 2011, \apjs, 194, 35

\bibitem[{{Den Hartog} {et~al.}(2014){Den Hartog}, {Ruffoni}, {Lawler}, {Pickering}, {Lind}, \& {Brewer}}]{denhartog2014}
{Den Hartog}, E.~A., {Ruffoni}, M.~P., {Lawler}, J.~E., {et~al.} 2014, \apjs, 215, 23

\bibitem[{{Drout} {et~al.}(2017){Drout}, {Piro}, {Shappee}, {Kilpatrick}, {Simon}, {Contreras}, {Coulter}, {Foley}, {Siebert}, {Morrell}, {Boutsia}, {Di Mille}, {Holoien}, {Kasen}, {Kollmeier}, {Madore}, {Monson}, {Murguia-Berthier}, {Pan}, {Prochaska}, {Ramirez-Ruiz}, {Rest}, {Adams}, {Alatalo}, {Ba{\~n}ados}, {Baughman}, {Beers}, {Bernstein}, {Bitsakis}, {Campillay}, {Hansen}, {Higgs}, {Ji}, {Maravelias}, {Marshall}, {Bidin}, {Prieto}, {Rasmussen}, {Rojas-Bravo}, {Strom}, {Ulloa}, {Vargas-Gonz{\'a}lez}, {Wan}, \& {Whitten}}]{drout2017}
{Drout}, M.~R., {Piro}, A.~L., {Shappee}, B.~J., {et~al.} 2017, Science, 358, 1570

\bibitem[{{Eichler} {et~al.}(2015){Eichler}, {Arcones}, {Kelic}, {Korobkin}, {Langanke}, {Marketin}, {Martinez-Pinedo}, {Panov}, {Rauscher}, {Rosswog}, {Winteler}, {Zinner}, \& {Thielemann}}]{eichler2015}
{Eichler}, M., {Arcones}, A., {Kelic}, A., {et~al.} 2015, \apj, 808, 30

\bibitem[{{Eitner} {et~al.}(2023){Eitner}, {Bergemann}, {Ruiter}, {Avril}, {Seitenzahl}, {Gent}, \& {C{\^o}t{\'e}}}]{eitner2023}
{Eitner}, P., {Bergemann}, M., {Ruiter}, A.~J., {et~al.} 2023, \aap, 677, A151

\bibitem[{{Ernandes} {et~al.}(2023){Ernandes}, {Castro}, {Barbuy}, {Spite}, {Hill}, {Castilho}, \& {Evans}}]{ernandes2023}
{Ernandes}, H., {Castro}, M.~J., {Barbuy}, B., {et~al.} 2023, \mnras, 524, 656

\bibitem[{{Ezzeddine} {et~al.}(2020){Ezzeddine}, {Rasmussen}, {Frebel}, {Chiti}, {Hinojisa}, {Placco}, {Ji}, {Beers}, {Hansen}, {Roederer}, {Sakari}, \& {Melendez}}]{ezzeddine2020}
{Ezzeddine}, R., {Rasmussen}, K., {Frebel}, A., {et~al.} 2020, \apj, 898, 150

\bibitem[{{Fedchak} \& {Lawler}(1999)}]{fedchak1999}
{Fedchak}, J.~A. \& {Lawler}, J.~E. 1999, \apj, 523, 734

\bibitem[{{Fitzpatrick} {et~al.}(2024){Fitzpatrick}, {Placco}, {Bolton}, {Merino}, {Ridgway}, \& {Stanghellini}}]{fitzpatrick2024}
{Fitzpatrick}, M., {Placco}, V., {Bolton}, A., {et~al.} 2024, arXiv e-prints, arXiv:2401.01982

\bibitem[{{Fivet} {et~al.}(2006){Fivet}, {Quinet}, {Bi{\'e}mont}, \& {Xu}}]{fivet2006}
{Fivet}, V., {Quinet}, P., {Bi{\'e}mont}, {\'E}., \& {Xu}, H.~L. 2006, Journal of Physics B Atomic Molecular Physics, 39, 3587

\bibitem[{{Fujimoto} {et~al.}(2008){Fujimoto}, {Nishimura}, \& {Hashimoto}}]{fujimoto2008}
{Fujimoto}, S.-i., {Nishimura}, N., \& {Hashimoto}, M.-a. 2008, \apj, 680, 1350

\bibitem[{{Fulbright}(2000)}]{fulbright2000}
{Fulbright}, J.~P. 2000, \aj, 120, 1841

\bibitem[{{Gaia Collaboration}(2018)}]{gaiarv}
{Gaia Collaboration}. 2018, VizieR Online Data Catalog, I/345

\bibitem[{{Gaia Collaboration} {et~al.}(2023){Gaia Collaboration}, {Vallenari}, {Brown}, {Prusti}, {de Bruijne}, {Arenou}, {Babusiaux}, {Biermann}, {Creevey}, {Ducourant}, {Evans}, {Eyer}, {Guerra}, {Hutton}, {Jordi}, {Klioner}, {Lammers}, {Lindegren}, {Luri}, {Mignard}, {Panem}, {Pourbaix}, {Randich}, {Sartoretti}, {Soubiran}, {Tanga}, {Walton}, {Bailer-Jones}, {Bastian}, {Drimmel}, {Jansen}, {Katz}, {Lattanzi}, {van Leeuwen}, {Bakker}, {Cacciari}, {Casta{\~n}eda}, {De Angeli}, {Fabricius}, {Fouesneau}, {Fr{\'e}mat}, {Galluccio}, {Guerrier}, {Heiter}, {Masana}, {Messineo}, {Mowlavi}, {Nicolas}, {Nienartowicz}, {Pailler}, {Panuzzo}, {Riclet}, {Roux}, {Seabroke}, {Sordo}, {Th{\'e}venin}, {Gracia-Abril}, {Portell}, {Teyssier}, {Altmann}, {Andrae}, {Audard}, {Bellas-Velidis}, {Benson}, {Berthier}, {Blomme}, {Burgess}, {Busonero}, {Busso}, {C{\'a}novas}, {Carry}, {Cellino}, {Cheek}, {Clementini}, {Damerdji}, {Davidson}, {de Teodoro}, {Nu{\~n}ez Campos}, {Delchambre}, {Dell'Oro}, {Esquej},
  {Fern{\'a}ndez-Hern{\'a}ndez}, {Fraile}, {Garabato}, {Garc{\'\i}a-Lario}, {Gosset}, {Haigron}, {Halbwachs}, {Hambly}, {Harrison}, {Hern{\'a}ndez}, {Hestroffer}, {Hodgkin}, {Holl}, {Jan{\ss}en}, {Jevardat de Fombelle}, {Jordan}, {Krone-Martins}, {Lanzafame}, {L{\"o}ffler}, {Marchal}, {Marrese}, {Moitinho}, {Muinonen}, {Osborne}, {Pancino}, {Pauwels}, {Recio-Blanco}, {Reyl{\'e}}, {Riello}, {Rimoldini}, {Roegiers}, {Rybizki}, {Sarro}, {Siopis}, {Smith}, {Sozzetti}, {Utrilla}, {van Leeuwen}, {Abbas}, {{\'A}brah{\'a}m}, {Abreu Aramburu}, {Aerts}, {Aguado}, {Ajaj}, {Aldea-Montero}, {Altavilla}, {{\'A}lvarez}, {Alves}, {Anders}, {Anderson}, {Anglada Varela}, {Antoja}, {Baines}, {Baker}, {Balaguer-N{\'u}{\~n}ez}, {Balbinot}, {Balog}, {Barache}, {Barbato}, {Barros}, {Barstow}, {Bartolom{\'e}}, {Bassilana}, {Bauchet}, {Becciani}, {Bellazzini}, {Berihuete}, {Bernet}, {Bertone}, {Bianchi}, {Binnenfeld}, {Blanco-Cuaresma}, {Blazere}, {Boch}, {Bombrun}, {Bossini}, {Bouquillon}, {Bragaglia}, {Bramante}, {Breedt},
  {Bressan}, {Brouillet}, {Brugaletta}, {Bucciarelli}, {Burlacu}, {Butkevich}, {Buzzi}, {Caffau}, {Cancelliere}, {Cantat-Gaudin}, {Carballo}, {Carlucci}, {Carnerero}, {Carrasco}, {Casamiquela}, {Castellani}, {Castro-Ginard}, {Chaoul}, {Charlot}, {Chemin}, {Chiaramida}, {Chiavassa}, {Chornay}, {Comoretto}, {Contursi}, {Cooper}, {Cornez}, {Cowell}, {Crifo}, {Cropper}, {Crosta}, {Crowley}, {Dafonte}, {Dapergolas}, {David}, {David}, {de Laverny}, {De Luise}, {De March}, {De Ridder}, {de Souza}, {de Torres}, {del Peloso}, {del Pozo}, {Delbo}, {Delgado}, {Delisle}, {Demouchy}, {Dharmawardena}, {Di Matteo}, {Diakite}, {Diener}, {Distefano}, {Dolding}, {Edvardsson}, {Enke}, {Fabre}, {Fabrizio}, {Faigler}, {Fedorets}, {Fernique}, {Fienga}, {Figueras}, {Fournier}, {Fouron}, {Fragkoudi}, {Gai}, {Garcia-Gutierrez}, {Garcia-Reinaldos}, {Garc{\'\i}a-Torres}, {Garofalo}, {Gavel}, {Gavras}, {Gerlach}, {Geyer}, {Giacobbe}, {Gilmore}, {Girona}, {Giuffrida}, {Gomel}, {Gomez}, {Gonz{\'a}lez-N{\'u}{\~n}ez},
  {Gonz{\'a}lez-Santamar{\'\i}a}, {Gonz{\'a}lez-Vidal}, {Granvik}, {Guillout}, {Guiraud}, {Guti{\'e}rrez-S{\'a}nchez}, {Guy}, {Hatzidimitriou}, {Hauser}, {Haywood}, {Helmer}, {Helmi}, {Sarmiento}, {Hidalgo}, {Hilger}, {H{\l}adczuk}, {Hobbs}, {Holland}, {Huckle}, {Jardine}, {Jasniewicz}, {Jean-Antoine Piccolo}, {Jim{\'e}nez-Arranz}, {Jorissen}, {Juaristi Campillo}, {Julbe}, {Karbevska}, {Kervella}, {Khanna}, {Kontizas}, {Kordopatis}, {Korn}, {K{\'o}sp{\'a}l}, {Kostrzewa-Rutkowska}, {Kruszy{\'n}ska}, {Kun}, {Laizeau}, {Lambert}, {Lanza}, {Lasne}, {Le Campion}, {Lebreton}, {Lebzelter}, {Leccia}, {Leclerc}, {Lecoeur-Taibi}, {Liao}, {Licata}, {Lindstr{\o}m}, {Lister}, {Livanou}, {Lobel}, {Lorca}, {Loup}, {Madrero Pardo}, {Magdaleno Romeo}, {Managau}, {Mann}, {Manteiga}, {Marchant}, {Marconi}, {Marcos}, {Marcos Santos}, {Mar{\'\i}n Pina}, {Marinoni}, {Marocco}, {Marshall}, {Martin Polo}, {Mart{\'\i}n-Fleitas}, {Marton}, {Mary}, {Masip}, {Massari}, {Mastrobuono-Battisti}, {Mazeh}, {McMillan}, {Messina}, {Michalik},
  {Millar}, {Mints}, {Molina}, {Molinaro}, {Moln{\'a}r}, {Monari}, {Mongui{\'o}}, {Montegriffo}, {Montero}, {Mor}, {Mora}, {Morbidelli}, {Morel}, {Morris}, {Muraveva}, {Murphy}, {Musella}, {Nagy}, {Noval}, {Oca{\~n}a}, {Ogden}, {Ordenovic}, {Osinde}, {Pagani}, {Pagano}, {Palaversa}, {Palicio}, {Pallas-Quintela}, {Panahi}, {Payne-Wardenaar}, {Pe{\~n}alosa Esteller}, {Penttil{\"a}}, {Pichon}, {Piersimoni}, {Pineau}, {Plachy}, {Plum}, {Poggio}, {Pr{\v{s}}a}, {Pulone}, {Racero}, {Ragaini}, {Rainer}, {Raiteri}, {Rambaux}, {Ramos}, {Ramos-Lerate}, {Re Fiorentin}, {Regibo}, {Richards}, {Rios Diaz}, {Ripepi}, {Riva}, {Rix}, {Rixon}, {Robichon}, {Robin}, {Robin}, {Roelens}, {Rogues}, {Rohrbasser}, {Romero-G{\'o}mez}, {Rowell}, {Royer}, {Ruz Mieres}, {Rybicki}, {Sadowski}, {S{\'a}ez N{\'u}{\~n}ez}, {Sagrist{\`a} Sell{\'e}s}, {Sahlmann}, {Salguero}, {Samaras}, {Sanchez Gimenez}, {Sanna}, {Santove{\~n}a}, {Sarasso}, {Schultheis}, {Sciacca}, {Segol}, {Segovia}, {S{\'e}gransan}, {Semeux}, {Shahaf}, {Siddiqui}, {Siebert},
  {Siltala}, {Silvelo}, {Slezak}, {Slezak}, {Smart}, {Snaith}, {Solano}, {Solitro}, {Souami}, {Souchay}, {Spagna}, {Spina}, {Spoto}, {Steele}, {Steidelm{\"u}ller}, {Stephenson}, {S{\"u}veges}, {Surdej}, {Szabados}, {Szegedi-Elek}, {Taris}, {Taylor}, {Teixeira}, {Tolomei}, {Tonello}, {Torra}, {Torra}, {Torralba Elipe}, {Trabucchi}, {Tsounis}, {Turon}, {Ulla}, {Unger}, {Vaillant}, {van Dillen}, {van Reeven}, {Vanel}, {Vecchiato}, {Viala}, {Vicente}, {Voutsinas}, {Weiler}, {Wevers}, {Wyrzykowski}, {Yoldas}, {Yvard}, {Zhao}, {Zorec}, {Zucker}, \& {Zwitter}}]{gaiadr3}
{Gaia Collaboration}, {Vallenari}, A., {Brown}, A.~G.~A., {et~al.} 2023, \aap, 674, A1

\bibitem[{{Grichener} \& {Soker}(2019)}]{grichener2019}
{Grichener}, A. \& {Soker}, N. 2019, \apj, 878, 24

\bibitem[{{Gudin} {et~al.}(2021){Gudin}, {Shank}, {Beers}, {Yuan}, {Limberg}, {Roederer}, {Placco}, {Holmbeck}, {Dietz}, {Rasmussen}, {Hansen}, {Sakari}, {Ezzeddine}, \& {Frebel}}]{gudin2021}
{Gudin}, D., {Shank}, D., {Beers}, T.~C., {et~al.} 2021, \apj, 908, 79

\bibitem[{{Gull} {et~al.}(2018){Gull}, {Frebel}, {Cain}, {Placco}, {Ji}, {Abate}, {Ezzeddine}, {Karakas}, {Hansen}, {Sakari}, {Holmbeck}, {Santucci}, {Casey}, \& {Beers}}]{gull2018}
{Gull}, M., {Frebel}, A., {Cain}, M.~G., {et~al.} 2018, \apj, 862, 174

\bibitem[{{Gurell} {et~al.}(2010){Gurell}, {Nilsson}, {Engstr{\"o}m}, {Lundberg}, {Blackwell-Whitehead}, {Nielsen}, \& {Mannervik}}]{gurell2010}
{Gurell}, J., {Nilsson}, H., {Engstr{\"o}m}, L., {et~al.} 2010, \aap, 511, A68

\bibitem[{{Hannaford} {et~al.}(1981){Hannaford}, {Larkins}, \& {Lowe}}]{hannaford1981}
{Hannaford}, P., {Larkins}, P.~L., \& {Lowe}, R.~M. 1981, Journal of Physics B Atomic Molecular Physics, 14, 2321

\bibitem[{{Hansen} {et~al.}(2019){Hansen}, {Beers}, {Ezzeddine}, {Frebel}, {Holmbeck}, {Ji}, {Marshall}, {Placco}, {Roederer}, \& {Sakari}}]{hansen2019}
{Hansen}, T.~T., {Beers}, T.~C., {Ezzeddine}, R., {et~al.} 2019, {Testing r-process nucleosynthesis models with two r-process enhanced stars}, HST Proposal. Cycle 27, ID. \#15951

\bibitem[{{Hansen} {et~al.}(2018){Hansen}, {Holmbeck}, {Beers}, {Placco}, {Roederer}, {Frebel}, {Sakari}, {Simon}, \& {Thompson}}]{hansen2018}
{Hansen}, T.~T., {Holmbeck}, E.~M., {Beers}, T.~C., {et~al.} 2018, \apj, 858, 92

\bibitem[{{Hattori} {et~al.}(2023){Hattori}, {Okuno}, \& {Roederer}}]{hattori2023}
{Hattori}, K., {Okuno}, A., \& {Roederer}, I.~U. 2023, \apj, 946, 48

\bibitem[{{Helmi} \& {de Zeeuw}(2000)}]{helmi2000}
{Helmi}, A. \& {de Zeeuw}, P.~T. 2000, \mnras, 319, 657

\bibitem[{{Holmbeck} {et~al.}(2018){Holmbeck}, {Beers}, {Roederer}, {Placco}, {Hansen}, {Sakari}, {Sneden}, {Liu}, {Lee}, {Cowan}, \& {Frebel}}]{holmbeck2018}
{Holmbeck}, E.~M., {Beers}, T.~C., {Roederer}, I.~U., {et~al.} 2018, \apjl, 859, L24

\bibitem[{{Holmbeck} {et~al.}(2020){Holmbeck}, {Hansen}, {Beers}, {Placco}, {Whitten}, {Rasmussen}, {Roederer}, {Ezzeddine}, {Sakari}, {Frebel}, {Drout}, {Simon}, {Thompson}, {Bland-Hawthorn}, {Gibson}, {Grebel}, {Kordopatis}, {Kunder}, {Mel{\'e}ndez}, {Navarro}, {Reid}, {Seabroke}, {Steinmetz}, {Watson}, \& {Wyse}}]{holmbeck2020}
{Holmbeck}, E.~M., {Hansen}, T.~T., {Beers}, T.~C., {et~al.} 2020, \apjs, 249, 30

\bibitem[{{Hunter}(2007)}]{matplotlib}
{Hunter}, J.~D. 2007, Computing in Science and Engineering, 9, 90

\bibitem[{{Ivans} {et~al.}(2006){Ivans}, {Simmerer}, {Sneden}, {Lawler}, {Cowan}, {Gallino}, \& {Bisterzo}}]{ivans2006}
{Ivans}, I.~I., {Simmerer}, J., {Sneden}, C., {et~al.} 2006, \apj, 645, 613

\bibitem[{{Ji} {et~al.}(2020){Ji}, {Li}, {Hansen}, {Casey}, {Koposov}, {Pace}, {Mackey}, {Lewis}, {Simpson}, {Bland-Hawthorn}, {Cullinane}, {Da Costa}, {Hattori}, {Martell}, {Kuehn}, {Erkal}, {Shipp}, {Wan}, \& {Zucker}}]{ji2020b}
{Ji}, A.~P., {Li}, T.~S., {Hansen}, T.~T., {et~al.} 2020, \aj, 160, 181

\bibitem[{{Kelson}(2003)}]{kelson2003}
{Kelson}, D.~D. 2003, \pasp, 115, 688

\bibitem[{{Kimble} {et~al.}(1998){Kimble}, {Woodgate}, {Bowers}, {Kraemer}, {Kaiser}, {Gull}, {Heap}, {Danks}, {Boggess}, {Green}, {Hutchings}, {Jenkins}, {Joseph}, {Linsky}, {Maran}, {Moos}, {Roesler}, {Timothy}, {Weistrop}, {Grady}, {Loiacono}, {Brown}, {Brumfield}, {Content}, {Feinberg}, {Isaacs}, {Krebs}, {Krueger}, {Melcher}, {Rebar}, {Vitagliano}, {Yagelowich}, {Meyer}, {Hood}, {Argabright}, {Becker}, {Bottema}, {Breyer}, {Bybee}, {Christon}, {Delamere}, {Dorn}, {Downey}, {Driggers}, {Ebbets}, {Gallegos}, {Garner}, {Hetlinger}, {Lettieri}, {Ludtke}, {Michika}, {Nyquist}, {Rose}, {Stocker}, {Sullivan}, {Van Houten}, {Woodruff}, {Baum}, {Hartig}, {Balzano}, {Biagetti}, {Blades}, {Bohlin}, {Clampin}, {Doxsey}, {Ferguson}, {Goudfrooij}, {Hulbert}, {Kutina}, {McGrath}, {Lindler}, {Beck}, {Feggans}, {Plait}, {Sandoval}, {Hill}, {Collins}, {Cornett}, {Fowler}, {Hill}, {Landsman}, {Malumuth}, {Standley}, {Blouke}, {Grusczak}, {Reed}, {Robinson}, {Valenti}, \& {Wolfe}}]{kimble1998}
{Kimble}, R.~A., {Woodgate}, B.~E., {Bowers}, C.~W., {et~al.} 1998, \apjl, 492, L83

\bibitem[{{Kramida} {et~al.}(2018){Kramida}, {Ralchenko}, {Nave}, \& {Reader}}]{kramida2018}
{Kramida}, A., {Ralchenko}, Y., {Nave}, G., \& {Reader}, J. 2018, in APS Meeting Abstracts, Vol. 2018, APS Division of Atomic, Molecular and Optical Physics Meeting Abstracts, M01.004

\bibitem[{{Kramida} {et~al.}(2020){Kramida}, {Ralchenko}, {Reader}, \& {NIST ASD Team}}]{kramida2020}
{Kramida}, A., {Ralchenko}, Y., {Reader}, J., \& {NIST ASD Team}. 2020, National Institute of Standards and Technology, Gaithersburg, MD.

\bibitem[{{Kurucz}(2011)}]{kurucz2011}
{Kurucz}, R.~L. 2011, Canadian Journal of Physics, 89, 417

\bibitem[{{Lattimer} \& {Schramm}(1974)}]{lattimer1974}
{Lattimer}, J.~M. \& {Schramm}, D.~N. 1974, \apjl, 192, L145

\bibitem[{{Lawler} {et~al.}(2001{\natexlab{a}}){Lawler}, {Bonvallet}, \& {Sneden}}]{lawler2001a}
{Lawler}, J.~E., {Bonvallet}, G., \& {Sneden}, C. 2001{\natexlab{a}}, \apj, 556, 452

\bibitem[{{Lawler} \& {Dakin}(1989)}]{lawler1989}
{Lawler}, J.~E. \& {Dakin}, J.~T. 1989, Journal of the Optical Society of America B Optical Physics, 6, 1457

\bibitem[{{Lawler} {et~al.}(2007){Lawler}, {den Hartog}, {Labby}, {Sneden}, {Cowan}, \& {Ivans}}]{lawler2007}
{Lawler}, J.~E., {den Hartog}, E.~A., {Labby}, Z.~E., {et~al.} 2007, \apjs, 169, 120

\bibitem[{{Lawler} {et~al.}(2006){Lawler}, {Den Hartog}, {Sneden}, \& {Cowan}}]{lawler2006}
{Lawler}, J.~E., {Den Hartog}, E.~A., {Sneden}, C., \& {Cowan}, J.~J. 2006, \apjs, 162, 227

\bibitem[{{Lawler} {et~al.}(2018){Lawler}, {Feigenson}, {Sneden}, {Cowan}, \& {Nave}}]{lawler2018}
{Lawler}, J.~E., {Feigenson}, T., {Sneden}, C., {Cowan}, J.~J., \& {Nave}, G. 2018, \apjs, 238, 7

\bibitem[{{Lawler} {et~al.}(2013){Lawler}, {Guzman}, {Wood}, {Sneden}, \& {Cowan}}]{lawler2013}
{Lawler}, J.~E., {Guzman}, A., {Wood}, M.~P., {Sneden}, C., \& {Cowan}, J.~J. 2013, \apjs, 205, 11

\bibitem[{{Lawler} {et~al.}(2019){Lawler}, {Hala}, {Sneden}, {Nave}, {Wood}, \& {Cowan}}]{lawler2019}
{Lawler}, J.~E., {Hala}, {Sneden}, C., {et~al.} 2019, \apjs, 241, 21

\bibitem[{{Lawler} {et~al.}(2004){Lawler}, {Sneden}, \& {Cowan}}]{lawler2004}
{Lawler}, J.~E., {Sneden}, C., \& {Cowan}, J.~J. 2004, \apj, 604, 850

\bibitem[{{Lawler} {et~al.}(2015){Lawler}, {Sneden}, \& {Cowan}}]{lawler2015}
{Lawler}, J.~E., {Sneden}, C., \& {Cowan}, J.~J. 2015, \apjs, 220, 13

\bibitem[{{Lawler} {et~al.}(2009){Lawler}, {Sneden}, {Cowan}, {Ivans}, \& {Den Hartog}}]{lawler2009}
{Lawler}, J.~E., {Sneden}, C., {Cowan}, J.~J., {Ivans}, I.~I., \& {Den Hartog}, E.~A. 2009, \apjs, 182, 51

\bibitem[{{Lawler} {et~al.}(2008){Lawler}, {Sneden}, {Cowan}, {Wyart}, {Ivans}, {Sobeck}, {Stockett}, \& {Den Hartog}}]{lawler2008}
{Lawler}, J.~E., {Sneden}, C., {Cowan}, J.~J., {et~al.} 2008, \apjs, 178, 71

\bibitem[{{Lawler} {et~al.}(2017){Lawler}, {Sneden}, {Nave}, {Den Hartog}, {Emraho{\u{g}}lu}, \& {Cowan}}]{lawler2017}
{Lawler}, J.~E., {Sneden}, C., {Nave}, G., {et~al.} 2017, \apjs, 228, 10

\bibitem[{{Lawler} {et~al.}(2001{\natexlab{b}}){Lawler}, {Wickliffe}, {Cowley}, \& {Sneden}}]{lawler2001b}
{Lawler}, J.~E., {Wickliffe}, M.~E., {Cowley}, C.~R., \& {Sneden}, C. 2001{\natexlab{b}}, \apjs, 137, 341

\bibitem[{{Lawler} {et~al.}(2001{\natexlab{c}}){Lawler}, {Wickliffe}, {den Hartog}, \& {Sneden}}]{lawler2001c}
{Lawler}, J.~E., {Wickliffe}, M.~E., {den Hartog}, E.~A., \& {Sneden}, C. 2001{\natexlab{c}}, \apj, 563, 1075

\bibitem[{{Lawler} {et~al.}(2001{\natexlab{d}}){Lawler}, {Wyart}, \& {Blaise}}]{lawler2001d}
{Lawler}, J.~E., {Wyart}, J.~F., \& {Blaise}, J. 2001{\natexlab{d}}, \apjs, 137, 351

\bibitem[{{Li} {et~al.}(2007){Li}, {Chatelain}, {Holt}, {Rehse}, {Rosner}, \& {Scholl}}]{li2007}
{Li}, R., {Chatelain}, R., {Holt}, R.~A., {et~al.} 2007, \physscr, 76, 577

\bibitem[{{Lind} {et~al.}(2011){Lind}, {Asplund}, {Barklem}, \& {Belyaev}}]{lind2011}
{Lind}, K., {Asplund}, M., {Barklem}, P.~S., \& {Belyaev}, A.~K. 2011, \aap, 528, A103

\bibitem[{{Lind} {et~al.}(2022){Lind}, {Nordlander}, {Wehrhahn}, {Montelius}, {Osorio}, {Barklem}, {Af{\c{s}}ar}, {Sneden}, \& {Kobayashi}}]{lind2022}
{Lind}, K., {Nordlander}, T., {Wehrhahn}, A., {et~al.} 2022, \aap, 665, A33

\bibitem[{{Ljung} {et~al.}(2006){Ljung}, {Nilsson}, {Asplund}, \& {Johansson}}]{ljung2006}
{Ljung}, G., {Nilsson}, H., {Asplund}, M., \& {Johansson}, S. 2006, \aap, 456, 1181

\bibitem[{{Lund} {et~al.}(2023){Lund}, {Engel}, {McLaughlin}, {Mumpower}, {Ney}, \& {Surman}}]{lund2023}
{Lund}, K.~A., {Engel}, J., {McLaughlin}, G.~C., {et~al.} 2023, \apj, 944, 144

\bibitem[{{Malcheva} {et~al.}(2006){Malcheva}, {Blagoev}, {Mayo}, {Ortiz}, {Xu}, {Svanberg}, {Quinet}, \& {Bi{\'e}mont}}]{malcheva2006}
{Malcheva}, G., {Blagoev}, K., {Mayo}, R., {et~al.} 2006, \mnras, 367, 754

\bibitem[{{Mallinson} {et~al.}(2022){Mallinson}, {Lind}, {Amarsi}, {Barklem}, {Grumer}, {Belyaev}, \& {Youakim}}]{mallinson2022}
{Mallinson}, J.~W.~E., {Lind}, K., {Amarsi}, A.~M., {et~al.} 2022, \aap, 668, A103

\bibitem[{{McCall}(2004)}]{mccall2004}
{McCall}, M.~L. 2004, \aj, 128, 2144

\bibitem[{{McMillan}(2017)}]{mcmillan2017}
{McMillan}, P.~J. 2017, \mnras, 465, 76

\bibitem[{{McWilliam}(1998)}]{mcwilliam1998}
{McWilliam}, A. 1998, \aj, 115, 1640

\bibitem[{{Mel{\'e}ndez} \& {Barbuy}(2009)}]{melendez2009}
{Mel{\'e}ndez}, J. \& {Barbuy}, B. 2009, \aap, 497, 611

\bibitem[{{Mucciarelli} {et~al.}(2021){Mucciarelli}, {Bellazzini}, \& {Massari}}]{mucciarelli2021}
{Mucciarelli}, A., {Bellazzini}, M., \& {Massari}, D. 2021, \aap, 653, A90

\bibitem[{{Munari} {et~al.}(2014){Munari}, {Henden}, {Frigo}, {Zwitter}, {Bienaym{\'e}}, {Bland-Hawthorn}, {Boeche}, {Freeman}, {Gibson}, {Gilmore}, {Grebel}, {Helmi}, {Kordopatis}, {Levine}, {Navarro}, {Parker}, {Reid}, {Seabroke}, {Siebert}, {Siviero}, {Smith}, {Steinmetz}, {Templeton}, {Terrell}, {Welch}, {Williams}, \& {Wyse}}]{munari2014}
{Munari}, U., {Henden}, A., {Frigo}, A., {et~al.} 2014, \aj, 148, 81

\bibitem[{{Nilsson} {et~al.}(2010){Nilsson}, {Hartman}, {Engstr{\"o}m}, {Lundberg}, {Sneden}, {Fivet}, {Palmeri}, {Quinet}, \& {Bi{\'e}mont}}]{nilsson2010}
{Nilsson}, H., {Hartman}, H., {Engstr{\"o}m}, L., {et~al.} 2010, \aap, 511, A16

\bibitem[{{Nilsson} {et~al.}(2006){Nilsson}, {Ljung}, {Lundberg}, \& {Nielsen}}]{nilsson2006}
{Nilsson}, H., {Ljung}, G., {Lundberg}, H., \& {Nielsen}, K.~E. 2006, \aap, 445, 1165

\bibitem[{{Nilsson} {et~al.}(2002){Nilsson}, {Zhang}, {Lundberg}, {Johansson}, \& {Nordstr{\"o}m}}]{nilsson2002}
{Nilsson}, H., {Zhang}, Z.~G., {Lundberg}, H., {Johansson}, S., \& {Nordstr{\"o}m}, B. 2002, \aap, 382, 368

\bibitem[{{O'Brian} {et~al.}(1991){O'Brian}, {Wickliffe}, {Lawler}, {Whaling}, \& {Brault}}]{obrian1991}
{O'Brian}, T.~R., {Wickliffe}, M.~E., {Lawler}, J.~E., {Whaling}, W., \& {Brault}, J.~W. 1991, Journal of the Optical Society of America B Optical Physics, 8, 1185

\bibitem[{Okuno \& Hattori(2022)}]{okuno2022}
Okuno, A. \& Hattori, K. 2022, A Greedy and Optimistic Approach to Clustering with a Specified Uncertainty of Covariates

\bibitem[{{Pakhomov} {et~al.}(2019){Pakhomov}, {Ryabchikova}, \& {Piskunov}}]{pakhomov2019}
{Pakhomov}, Y.~V., {Ryabchikova}, T.~A., \& {Piskunov}, N.~E. 2019, Astronomy Reports, 63, 1010

\bibitem[{{Patel} {et~al.}(2025){Patel}, {Metzger}, {Goldberg}, {Cehula}, {Thompson}, \& {Renzo}}]{patel2025}
{Patel}, A., {Metzger}, B.~D., {Goldberg}, J.~A., {et~al.} 2025, arXiv e-prints, arXiv:2501.17253

\bibitem[{{Pehlivan Rhodin} {et~al.}(2017){Pehlivan Rhodin}, {Hartman}, {Nilsson}, \& {J{\"o}nsson}}]{pehlivan2017}
{Pehlivan Rhodin}, A., {Hartman}, H., {Nilsson}, H., \& {J{\"o}nsson}, P. 2017, \aap, 598, A102

\bibitem[{{Peterson} {et~al.}(2020){Peterson}, {Barbuy}, \& {Spite}}]{peterson2020}
{Peterson}, R.~C., {Barbuy}, B., \& {Spite}, M. 2020, \aap, 638, A64

\bibitem[{{Peterson} {et~al.}(2001){Peterson}, {Dorman}, \& {Rood}}]{peterson2001}
{Peterson}, R.~C., {Dorman}, B., \& {Rood}, R.~T. 2001, \apj, 559, 372

\bibitem[{{Pickering} {et~al.}(2001){Pickering}, {Thorne}, \& {Perez}}]{pickering2001}
{Pickering}, J.~C., {Thorne}, A.~P., \& {Perez}, R. 2001, \apjs, 132, 403

\bibitem[{{Pickering} {et~al.}(2002){Pickering}, {Thorne}, \& {Perez}}]{pickering2002}
{Pickering}, J.~C., {Thorne}, A.~P., \& {Perez}, R. 2002, \apjs, 138, 247

\bibitem[{{Piskunov} {et~al.}(1995){Piskunov}, {Kupka}, {Ryabchikova}, {Weiss}, \& {Jeffery}}]{piskunov1995}
{Piskunov}, N.~E., {Kupka}, F., {Ryabchikova}, T.~A., {Weiss}, W.~W., \& {Jeffery}, C.~S. 1995, \aaps, 112, 525

\bibitem[{{Placco} {et~al.}(2017){Placco}, {Holmbeck}, {Frebel}, {Beers}, {Surman}, {Ji}, {Ezzeddine}, {Points}, {Kaleida}, {Hansen}, {Sakari}, \& {Casey}}]{placco2017}
{Placco}, V.~M., {Holmbeck}, E.~M., {Frebel}, A., {et~al.} 2017, \apj, 844, 18

\bibitem[{{Placco} {et~al.}(2020){Placco}, {Santucci}, {Yuan}, {Mardini}, {Holmbeck}, {Wang}, {Surman}, {Hansen}, {Roederer}, {Beers}, {Choplin}, {Ji}, {Ezzeddine}, {Frebel}, {Sakari}, {Whitten}, \& {Zepeda}}]{placco2020}
{Placco}, V.~M., {Santucci}, R.~M., {Yuan}, Z., {et~al.} 2020, \apj, 897, 78

\bibitem[{{Placco} {et~al.}(2021){Placco}, {Sneden}, {Roederer}, {Lawler}, {Den Hartog}, {Hejazi}, {Maas}, \& {Bernath}}]{placco2021}
{Placco}, V.~M., {Sneden}, C., {Roederer}, I.~U., {et~al.} 2021, Research Notes of the American Astronomical Society, 5, 92

\bibitem[{{Reggiani} {et~al.}(2019){Reggiani}, {Amarsi}, {Lind}, {Barklem}, {Zatsarinny}, {Bartschat}, {Fursa}, {Bray}, {Spina}, \& {Mel{\'e}ndez}}]{reggiani2019}
{Reggiani}, H., {Amarsi}, A.~M., {Lind}, K., {et~al.} 2019, \aap, 627, A177

\bibitem[{{Roederer} {et~al.}(2024){Roederer}, {Beers}, {Hattori}, {Placco}, {Hansen}, {Ezzeddine}, {Frebel}, {Holmbeck}, \& {Sakari}}]{roederer2024}
{Roederer}, I.~U., {Beers}, T.~C., {Hattori}, K., {et~al.} 2024, \apj, 971, 158

\bibitem[{{Roederer} {et~al.}(2022{\natexlab{a}}){Roederer}, {Cowan}, {Pignatari}, {Beers}, {Den Hartog}, {Ezzeddine}, {Frebel}, {Hansen}, {Holmbeck}, {Mumpower}, {Placco}, {Sakari}, {Surman}, \& {Vassh}}]{roederer2022b}
{Roederer}, I.~U., {Cowan}, J.~J., {Pignatari}, M., {et~al.} 2022{\natexlab{a}}, \apj, 936, 84

\bibitem[{{Roederer} {et~al.}(2018{\natexlab{a}}){Roederer}, {Hattori}, \& {Valluri}}]{roederer2018b}
{Roederer}, I.~U., {Hattori}, K., \& {Valluri}, M. 2018{\natexlab{a}}, \aj, 156, 179

\bibitem[{{Roederer} \& {Lawler}(2012)}]{roedererlawler2012}
{Roederer}, I.~U. \& {Lawler}, J.~E. 2012, \apj, 750, 76

\bibitem[{{Roederer} \& {Lawler}(2021)}]{roederer2021}
{Roederer}, I.~U. \& {Lawler}, J.~E. 2021, \apj, 912, 119

\bibitem[{{Roederer} {et~al.}(2022{\natexlab{b}}){Roederer}, {Lawler}, {Den Hartog}, {Placco}, {Surman}, {Beers}, {Ezzeddine}, {Frebel}, {Hansen}, {Hattori}, {Holmbeck}, \& {Sakari}}]{roederer2022a}
{Roederer}, I.~U., {Lawler}, J.~E., {Den Hartog}, E.~A., {et~al.} 2022{\natexlab{b}}, \apjs, 260, 27

\bibitem[{{Roederer} {et~al.}(2008){Roederer}, {Lawler}, {Sneden}, {Cowan}, {Sobeck}, \& {Pilachowski}}]{roederer2008}
{Roederer}, I.~U., {Lawler}, J.~E., {Sneden}, C., {et~al.} 2008, \apj, 675, 723

\bibitem[{{Roederer} {et~al.}(2012){Roederer}, {Lawler}, {Sobeck}, {Beers}, {Cowan}, {Frebel}, {Ivans}, {Schatz}, {Sneden}, \& {Thompson}}]{roederer2012}
{Roederer}, I.~U., {Lawler}, J.~E., {Sobeck}, J.~S., {et~al.} 2012, \apjs, 203, 27

\bibitem[{{Roederer} {et~al.}(2014{\natexlab{a}}){Roederer}, {Preston}, {Thompson}, {Shectman}, {Sneden}, {Burley}, \& {Kelson}}]{roederer2014a}
{Roederer}, I.~U., {Preston}, G.~W., {Thompson}, I.~B., {et~al.} 2014{\natexlab{a}}, \aj, 147, 136

\bibitem[{{Roederer} {et~al.}(2018{\natexlab{b}}){Roederer}, {Sakari}, {Placco}, {Beers}, {Ezzeddine}, {Frebel}, \& {Hansen}}]{roederer2018a}
{Roederer}, I.~U., {Sakari}, C.~M., {Placco}, V.~M., {et~al.} 2018{\natexlab{b}}, \apj, 865, 129

\bibitem[{{Roederer} {et~al.}(2014{\natexlab{b}}){Roederer}, {Schatz}, {Lawler}, {Beers}, {Cowan}, {Frebel}, {Ivans}, {Sneden}, \& {Sobeck}}]{roederer2014b}
{Roederer}, I.~U., {Schatz}, H., {Lawler}, J.~E., {et~al.} 2014{\natexlab{b}}, \apj, 791, 32

\bibitem[{{Roederer} {et~al.}(2010){Roederer}, {Sneden}, {Lawler}, \& {Cowan}}]{roederer2010}
{Roederer}, I.~U., {Sneden}, C., {Lawler}, J.~E., \& {Cowan}, J.~J. 2010, \apjl, 714, L123

\bibitem[{{Roederer} {et~al.}(2023){Roederer}, {Vassh}, {Holmbeck}, {Mumpower}, {Surman}, {Cowan}, {Beers}, {Ezzeddine}, {Frebel}, {Hansen}, {Placco}, \& {Sakari}}]{roederer2023}
{Roederer}, I.~U., {Vassh}, N., {Holmbeck}, E.~M., {et~al.} 2023, Science, 382, 1177

\bibitem[{{Ruffoni} {et~al.}(2014){Ruffoni}, {Den Hartog}, {Lawler}, {Brewer}, {Lind}, {Nave}, \& {Pickering}}]{ruffoni2014}
{Ruffoni}, M.~P., {Den Hartog}, E.~A., {Lawler}, J.~E., {et~al.} 2014, \mnras, 441, 3127

\bibitem[{{Sakari} {et~al.}(2018{\natexlab{a}}){Sakari}, {Placco}, {Hansen}, {Holmbeck}, {Beers}, {Frebel}, {Roederer}, {Venn}, {Wallerstein}, {Davis}, {Farrell}, \& {Yong}}]{sakari2018b}
{Sakari}, C.~M., {Placco}, V.~M., {Hansen}, T., {et~al.} 2018{\natexlab{a}}, \apjl, 854, L20

\bibitem[{{Sakari} {et~al.}(2018{\natexlab{b}}){Sakari}, {Placco}, {Hansen}, {Holmbeck}, {Beers}, {Frebel}, {Roederer}, {Venn}, {Wallerstein}, {Davis}, {Farrell}, \& {Yong}}]{sakari2018a}
{Sakari}, C.~M., {Placco}, V.~M., {Hansen}, T., {et~al.} 2018{\natexlab{b}}, \apjl, 854, L20

\bibitem[{{Schlafly} \& {Finkbeiner}(2011)}]{schlafly2011}
{Schlafly}, E.~F. \& {Finkbeiner}, D.~P. 2011, \apj, 737, 103

\bibitem[{{Shah} {et~al.}(2024){Shah}, {Ezzeddine}, {Roederer}, {Hansen}, {Placco}, {Beers}, {Frebel}, {Ji}, {Holmbeck}, {Marshall}, \& {Sakari}}]{shah2024}
{Shah}, S.~P., {Ezzeddine}, R., {Roederer}, I.~U., {et~al.} 2024, \mnras, 529, 1917

\bibitem[{{Shank} {et~al.}(2023){Shank}, {Beers}, {Placco}, {Gudin}, {Catapano}, {Holmbeck}, {Ezzeddine}, {Roederer}, {Sakari}, {Frebel}, \& {Hansen}}]{shank2023}
{Shank}, D., {Beers}, T.~C., {Placco}, V.~M., {et~al.} 2023, \apj, 943, 23

\bibitem[{{Siegel} {et~al.}(2019){Siegel}, {Barnes}, \& {Metzger}}]{siegel2019}
{Siegel}, D.~M., {Barnes}, J., \& {Metzger}, B.~D. 2019, \nat, 569, 241

\bibitem[{{Sikstr{\"o}m} {et~al.}(2001){Sikstr{\"o}m}, {Pihlemark}, {Nilsson}, {Litz{\'e}n}, {Johansson}, {Li}, \& {Lundberg}}]{sikstrom2001}
{Sikstr{\"o}m}, C.~M., {Pihlemark}, H., {Nilsson}, H., {et~al.} 2001, Journal of Physics B Atomic Molecular Physics, 34, 477

\bibitem[{{Siqueira Mello} {et~al.}(2013){Siqueira Mello}, {Spite}, {Barbuy}, {Spite}, {Caffau}, {Hill}, {Wanajo}, {Primas}, {Plez}, {Cayrel}, {Andersen}, {Nordstr{\"o}m}, {Sneden}, {Beers}, {Bonifacio}, {Fran{\c{c}}ois}, \& {Molaro}}]{siqueira2013}
{Siqueira Mello}, C., {Spite}, M., {Barbuy}, B., {et~al.} 2013, \aap, 550, A122

\bibitem[{{Sneden} {et~al.}(2008){Sneden}, {Cowan}, \& {Gallino}}]{sneden2008}
{Sneden}, C., {Cowan}, J.~J., \& {Gallino}, R. 2008, \araa, 46, 241

\bibitem[{{Sneden} {et~al.}(2003){Sneden}, {Cowan}, {Lawler}, {Ivans}, {Burles}, {Beers}, {Primas}, {Hill}, {Truran}, {Fuller}, {Pfeiffer}, \& {Kratz}}]{sneden2003}
{Sneden}, C., {Cowan}, J.~J., {Lawler}, J.~E., {et~al.} 2003, \apj, 591, 936

\bibitem[{{Sneden} {et~al.}(2009){Sneden}, {Lawler}, {Cowan}, {Ivans}, \& {Den Hartog}}]{sneden2009}
{Sneden}, C., {Lawler}, J.~E., {Cowan}, J.~J., {Ivans}, I.~I., \& {Den Hartog}, E.~A. 2009, \apjs, 182, 80

\bibitem[{{Sneden}(1973)}]{sneden1973}
{Sneden}, C.~A. 1973, PhD thesis, University of Texas, Austin

\bibitem[{{Sobeck} {et~al.}(2011){Sobeck}, {Kraft}, {Sneden}, {Preston}, {Cowan}, {Smith}, {Thompson}, {Shectman}, \& {Burley}}]{sobeck2011}
{Sobeck}, J.~S., {Kraft}, R.~P., {Sneden}, C., {et~al.} 2011, \aj, 141, 175

\bibitem[{{Sobeck} {et~al.}(2007){Sobeck}, {Lawler}, \& {Sneden}}]{sobeck2007}
{Sobeck}, J.~S., {Lawler}, J.~E., \& {Sneden}, C. 2007, \apj, 667, 1267

\bibitem[{{Surman} \& {Engel}(2001)}]{surman2001}
{Surman}, R. \& {Engel}, J. 2001, \prc, 64, 035801

\bibitem[{{Surman} {et~al.}(2008){Surman}, {McLaughlin}, {Ruffert}, {Janka}, \& {Hix}}]{surman2008}
{Surman}, R., {McLaughlin}, G.~C., {Ruffert}, M., {Janka}, H.~T., \& {Hix}, W.~R. 2008, \apjl, 679, L117

\bibitem[{{Tody}(1986)}]{tody1986}
{Tody}, D. 1986, in Society of Photo-Optical Instrumentation Engineers (SPIE) Conference Series, Vol. 627, Instrumentation in astronomy VI, ed. D.~L. {Crawford}, 733

\bibitem[{{Tody}(1993)}]{tody1993}
{Tody}, D. 1993, in Astronomical Society of the Pacific Conference Series, Vol.~52, Astronomical Data Analysis Software and Systems II, ed. R.~J. {Hanisch}, R.~J.~V. {Brissenden}, \& J.~{Barnes}, 173

\bibitem[{{Tr{\"a}bert} {et~al.}(1999){Tr{\"a}bert}, {Wolf}, {Linkemann}, \& {Tordoir}}]{trabert1999}
{Tr{\"a}bert}, E., {Wolf}, A., {Linkemann}, J., \& {Tordoir}, X. 1999, Journal of Physics B Atomic Molecular Physics, 32, 537

\bibitem[{{van der Walt} {et~al.}(2011){van der Walt}, {Colbert}, \& {Varoquaux}}]{numpy}
{van der Walt}, S., {Colbert}, S.~C., \& {Varoquaux}, G. 2011, Computing in Science and Engineering, 13, 22

\bibitem[{{Vasiliev}(2019)}]{vasiliev2019}
{Vasiliev}, E. 2019, \mnras, 482, 1525

\bibitem[{{Wehmeyer} {et~al.}(2019){Wehmeyer}, {Fr{\"o}hlich}, {C{\^o}t{\'e}}, {Pignatari}, \& {Thielemann}}]{wehmeyer2019}
{Wehmeyer}, B., {Fr{\"o}hlich}, C., {C{\^o}t{\'e}}, B., {Pignatari}, M., \& {Thielemann}, F.~K. 2019, \mnras, 487, 1745

\bibitem[{{Wenger} {et~al.}(2000){Wenger}, {Ochsenbein}, {Egret}, {Dubois}, {Bonnarel}, {Borde}, {Genova}, {Jasniewicz}, {Lalo{\"e}}, {Lesteven}, \& {Monier}}]{wenger2000}
{Wenger}, M., {Ochsenbein}, F., {Egret}, D., {et~al.} 2000, \aaps, 143, 9

\bibitem[{{Wickliffe} \& {Lawler}(1997)}]{wickliffe1997}
{Wickliffe}, M.~E. \& {Lawler}, J.~E. 1997, Journal of the Optical Society of America B Optical Physics, 14, 737

\bibitem[{{Wickliffe} {et~al.}(2000){Wickliffe}, {Lawler}, \& {Nave}}]{wickliffe2000}
{Wickliffe}, M.~E., {Lawler}, J.~E., \& {Nave}, G. 2000, \jqsrt, 66, 363

\bibitem[{{Wood} {et~al.}(2014{\natexlab{a}}){Wood}, {Lawler}, {Den Hartog}, {Sneden}, \& {Cowan}}]{wood2014a}
{Wood}, M.~P., {Lawler}, J.~E., {Den Hartog}, E.~A., {Sneden}, C., \& {Cowan}, J.~J. 2014{\natexlab{a}}, \apjs, 214, 18

\bibitem[{{Wood} {et~al.}(2013){Wood}, {Lawler}, {Sneden}, \& {Cowan}}]{wood2013}
{Wood}, M.~P., {Lawler}, J.~E., {Sneden}, C., \& {Cowan}, J.~J. 2013, \apjs, 208, 27

\bibitem[{{Wood} {et~al.}(2014{\natexlab{b}}){Wood}, {Lawler}, {Sneden}, \& {Cowan}}]{wood2014b}
{Wood}, M.~P., {Lawler}, J.~E., {Sneden}, C., \& {Cowan}, J.~J. 2014{\natexlab{b}}, \apjs, 211, 20

\bibitem[{{Woodgate} {et~al.}(1998){Woodgate}, {Kimble}, {Bowers}, {Kraemer}, {Kaiser}, {Danks}, {Grady}, {Loiacono}, {Brumfield}, {Feinberg}, {Gull}, {Heap}, {Maran}, {Lindler}, {Hood}, {Meyer}, {Vanhouten}, {Argabright}, {Franka}, {Bybee}, {Dorn}, {Bottema}, {Woodruff}, {Michika}, {Sullivan}, {Hetlinger}, {Ludtke}, {Stocker}, {Delamere}, {Rose}, {Becker}, {Garner}, {Timothy}, {Blouke}, {Joseph}, {Hartig}, {Green}, {Jenkins}, {Linsky}, {Hutchings}, {Moos}, {Boggess}, {Roesler}, \& {Weistrop}}]{woodgate1998}
{Woodgate}, B.~E., {Kimble}, R.~A., {Bowers}, C.~W., {et~al.} 1998, \pasp, 110, 1183

\bibitem[{{Xu} {et~al.}(2004){Xu}, {Persson}, {Svanberg}, {Blagoev}, {Malcheva}, {Pentchev}, {Bi{\'e}mont}, {Campos}, {Ortiz}, \& {Mayo}}]{xu2004}
{Xu}, H.~L., {Persson}, A., {Svanberg}, S., {et~al.} 2004, \pra, 70, 042508

\end{thebibliography}

\begin{appendix} 

\section{Atomic data}
\begin{table}[ht]
\caption{Atomic data and abundances for individual lines analysed.}
\label{tab:lines}
\resizebox{\columnwidth}{!}{
    \centering
    \begin{tabular}{llrrrrr}
    \hline\hline
$\lambda$  & species & $\chi$ & $\log$ $gf$ & EW & $\log\epsilon$ (X) & Ref\\
 (\Angstrom) & & (eV) &  & (m\Angstrom) & (dex) & \\    
\hline
7771.94 &  \ion{O}{1} &  9.15 &  $+$0.37 & 26.64  & $+$7.13 & 1 \\
7774.17 &  \ion{O}{1} &  9.15 &  $+$0.22 & 21.02  & $+$7.14 & 1 \\
7775.39 &  \ion{O}{1} &  9.15 &     0.00 & 13.73  & $+$7.12 & 1 \\ 
5889.95 & \ion{Na}{1} &  0.00 &  $+$0.11 &177.37  & $+$4.62 & 1 \\
5895.92 & \ion{Na}{1} &  0.00 &  $-$0.19 &150.18  & $+$4.46 & 1 \\
2776.69 & \ion{Mg}{1} &  2.71 &  $-$0.12 &   syn  & $+$5.39 & 1 \\
2779.83 & \ion{Mg}{1} &  2.72 &  $+$0.38 &   syn  & $+$5.37 & 1 \\
2782.97 & \ion{Mg}{1} &  2.72 &  $-$0.12 &   syn  & $+$5.43 & 1 \\
2846.72 & \ion{Mg}{1} &  2.71 &  $-$1.32 &   syn  & $+$5.31 & 1 \\
3829.35 & \ion{Mg}{1} &  2.71 &  $-$0.23 &166.60  & $+$5.60 & 1 \\
3986.75 & \ion{Mg}{1} &  4.35 &  $-$1.06 & 22.42  & $+$5.55 & 1 \\
4057.51 & \ion{Mg}{1} &  4.35 &  $-$0.90 & 23.82  & $+$5.42 & 1 \\
\\
\hline
    \end{tabular}}
     \tablebib{(1) NIST \cite{kramida2020}; (2) \citet{pehlivan2017}; (3) NIST \cite{kramida2020} for log($gf$) value and VALD \citet{piskunov1995,pakhomov2019} for HFS; (4) \citet{trabert1999} for log($gf$) value and \citet{roederer2021} for HFS;
(5) \citet{denhartog2023}; (6) \citet{denhartog2021b}; (7) \citet{lawler1989} for $\log g$ and \citet{kurucz2011} for HFS; (8) \citet{lawler2019};
(9) \citet{lawler2013}; (10) \citet{wood2013}; (11) \citet{pickering2001}, with corrections given in \citet{pickering2002}; (12) \citet{wood2014a} for log($gf$) value and HFS; (13) \citet{sobeck2007}, (14) \citet{lawler2017}, (15) \citet{gurell2010}; (16) \citet{nilsson2006}; (17) \citet{denhartog2011} for both log($gf$) value and hfs; (18) \citet{obrian1991}; (19) \citet{belmonte2017}; (20) \citet{denhartog2014}; (21) \citet{ruffoni2014}; (22) \citet{denhartog2019}; (23) \citet{melendez2009}; (24) \citet{lawler2015} for log($gf$) values and HFS; (25) \citet{lawler2018} for log($gf$) value and \citet{roederer2022a}for HFS; (26) \citet{wood2014b}; (27) \citet{fedchak1999}; (28) \citet{roedererlawler2012}; (29) \citet{biemont2011}; (30) \citet{ljung2006}; (31) \citet{malcheva2006}; (32) \citet{nilsson2010}, including HFS; (33) \citet{sikstrom2001}; (34) \citet{xu2004}; (35) \citet{kramida2018}, using HFS/IS from \citet{mcwilliam1998}; (36) \citet{lawler2001a}, using HFS from \citet{ivans2006} when available; (37) \citet{lawler2009}; (38) \citet{li2007}, using HFS from \citet{sneden2009}; (39) \citet{denhartog2003}, using HFS/IS from \citet{roederer2008} when available; (40) \citet{lawler2006}, using HFS/IS from \citet{roederer2008} when available; (41) \citet{lawler2001c}, using HFS/IS from \citet{ivans2006}; (42) \citet{denhartog2006}, (43) \citet{lawler2001b}, using HFS from \citet{lawler2001d}; (44) \citet{wickliffe2000}; (45) \citet{lawler2004}, using HFS from \citet{lawler2009}; (46) \citet{lawler2008}; (47) \citet{wickliffe1997}; (48) \citet{sneden2009} for log($gf$) value and HFS/IS; (49) \citet{roederer2010} for log($gf$) value and \citet{denhartog2020} for HFS; (50) \citet{denhartog2021a}; (51) \citet{lawler2007}; (52) \citet{denhartog2005} for log($gf$) value and HFS/IS; (54) \citet{denhartog2005} for log($gf$) value only; (55) \citet{hannaford1981} for log($gf$) value and \citet{demidov2021}for HFS; (56) \citet{nilsson2002}}
\tablefoot{The complete version of this table is available online only. A subset is shown here to illustrate its form and content.}    
\end{table}

\end{appendix}
\end{document}